\documentclass[review]{elsarticle}

\usepackage{lineno, hyperref}

\usepackage{graphics}

\usepackage{xcolor}
\usepackage{amssymb}
\usepackage{graphicx}
\usepackage{mathtools} 
\usepackage{multirow}
\usepackage{booktabs}
\usepackage{stackengine}
\usepackage{scalerel}
\usepackage{tikz}

\newcommand\openbigstar[1][0.7]{%
  \scalerel*{%
    \stackinset{c}{-.125pt}{c}{}{\scalebox{#1}{\color{white}{$\bigstar$}}}{%
      $\bigstar$}%
  }{\bigstar}
}

\newcommand{\bDiamond}{\mathbin{\Diamond}}
\newcommand{\bLozenge}{\mathbin{\blacklozenge}}

\journal{ International Journal of Multiphase Flows }

\biboptions{authoryear}

\begin{document}

\begin{frontmatter}

\title{Cavitating Flow Behind a Backward Facing Step}

\author{Anubhav Bhatt, Harish Ganesh, S.L. Ceccio }
\address{University Of Michigan, Ann Arbor}



\begin{abstract}

Flow topology and unsteady cavitation dynamics in the wake of a backward facing step was investigated using high speed videography and time-resolved X-Ray densitometry, along with static and dynamic pressure measurements. The measurements are used to inform the understanding of underlying mechanisms of observed flow dynamics as they are related to shock wave induced instabilities. The differences in cavity topology and behaviour at different cavitation numbers are examined to highlight flow features such as the pair of cavitating spanwise vortices in the shear layer and propagating bubbly shock front. The shock speeds at different cavitation conditions are estimated based on void-fraction measurements using the X-T diagram and compared to those computed using the one-dimensional Rankine Hugoniot jump condition. The two speeds(computed and measured) are found to be comparable. The effect of compressibility of the bubbly mixture is determined by estimating the local speed of sound using homogeneous `frozen model' and homogeneous `equilibrium model'. The current estimation of sound speed suggests that expression used to determine the local speed of sound is closer to the classic `frozen model' than the `equilibrium model' of speed of sound.          

\end{abstract}

\begin{keyword}
Shockwaves, speed of sound, X-Ray densitometry

\end{keyword}

\end{frontmatter}


\section{Introduction}




Developed cavitation in reattaching shear flows can result in unwanted noise, erosion and vibration various hydrodynamic applications. Despite being extensively studied and reported, the complexity of these turbulent, high void fraction flow makes direct interrogation quite challenging. Given the geometric simplicity and the availability of previous results detailing the single phase phenomena, the flow behind a backward facing step is an ideal candidate for experimental investigation of the effect of cavitation on turbulent shear layers.

\begin{figure}
	\centering
	\includegraphics[width=0.7\linewidth]{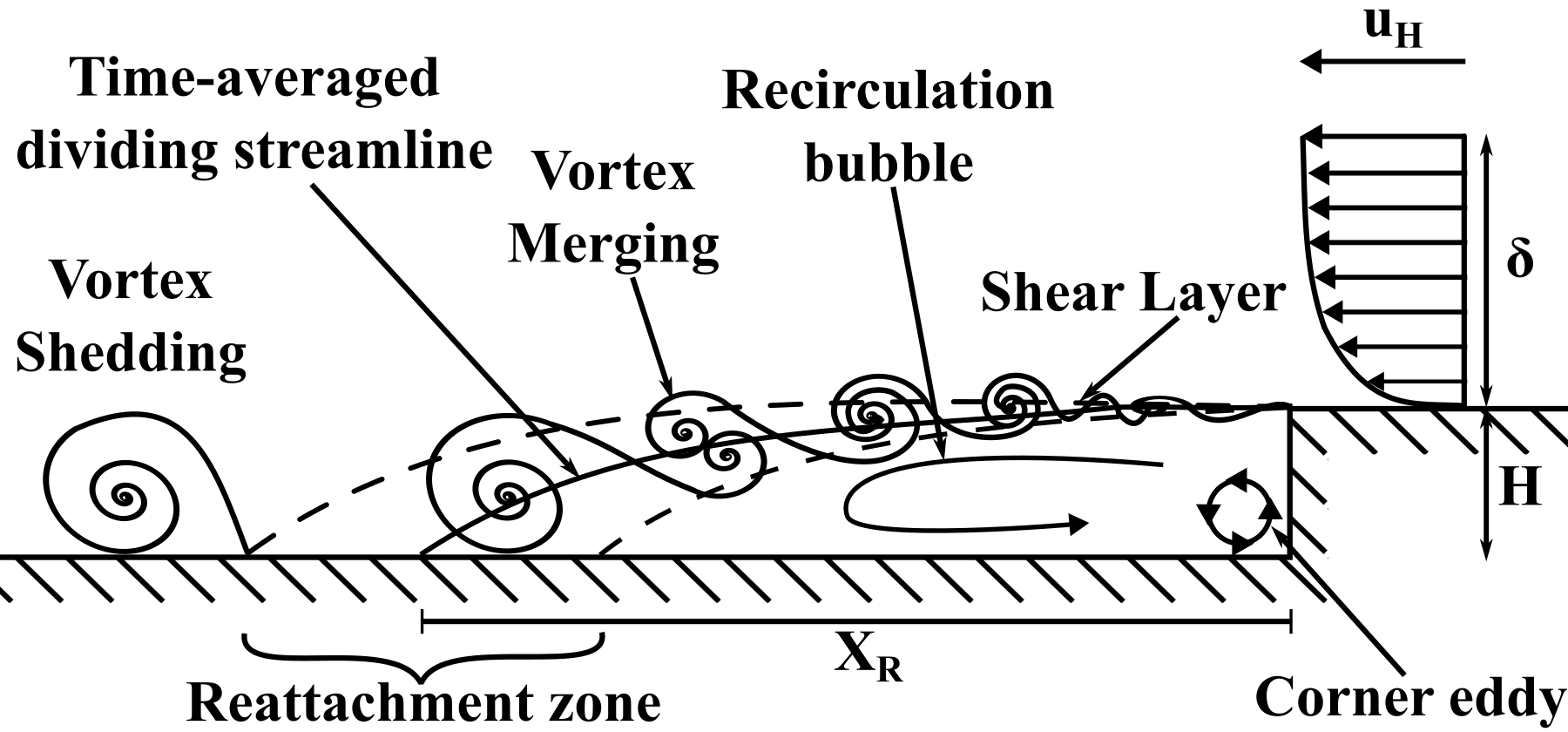}
	\caption{Topology of the non-cavitating flow behind a backward facing step.}
	\label{fig:figure1}
\end{figure}

A schematic of the single-phase flow topology downstream of the backward facing step is shown in Figure \ref{fig:figure1}.  The flow has a sharp separation point, fixed to the edge of the step. As the single phase flow separates from the edge, the free shear flow rolls up to create large scale spanwise vortices, similar to those of mixing free shear layer structures (\cite{troutt1984organized}). This shear flow will ultimately reattach to form a recirculation bubble underneath the shear layer. Further downstream the wall effects start to dominate (\cite{bradshaw1972reattachment}) as shear layer curves sharply downward in the reattachment region and impinges on the wall. In the reattachment region, the shear layer is subjected to distorting forces due to effects of stabilizing curvature, adverse pressure gradient and strong interaction with the wall (\cite{eaton1981review}). Downstream of the reattachment a new sub-boundary layer begins to form. It can be clearly seen that this geometrically simple flow has almost all the features of separating and reattaching shear flows including a mixing layer, recirculation, re-attachment and a redeveloping boundary layer. In addition, for backward facing step flow configurations, the boundary of the flow above the step may be in close proximity to the step to influence the step flow, making the expansion ratio also a parameter of interest (\cite{biswas2004backward}). 

 \cite{bernal1986streamwise} developed a plane mixing-layer model highlighting the streamwise vortex structures and noted structures of dominant spanwise vortices.They reported the ‘braid’ region between adjacent spanwise vortices to contain an array of streamwise vortices which snake between neighboring eddies. The spanwise vortex structures shed near the separation point can be subjected to vortex pairing interaction upstream of reattachment region, indicating that the growth of the separated layer may be strongly dependent on the pairing process (\cite{troutt1984organized}). Using PIV measurements and pattern recognition \cite{scarano1999pattern} presented details of the flow structures observed within the shear layer. 

From a dynamics perspective, a low frequency “flapping motion” in the whole separated region has been observed (\cite{eaton1981review}). One possible explanation for this motion is attributed to the mass imbalance between shear layer entrainment from recirculation region and re-entrant flow near the reattachment point(\cite{eaton1981review}). This imbalance is said to be caused by short term breakdown of spanwise vortices in shear layer that leads to decrease in the entrainment rate, causing a change in volume of recirculating fluid. Another explanation is attributed to a reduction in the amount of mass engulfed by the separation bubble (\cite{driver1987time}) causing the bubble to momentarily collapse. Yet another explanation of flapping motion states that the re-attachment point is unsteady and it moves upstream intermittently in bursts, that splits shear layer into two halves and compresses the fluid beneath the shear layer (\cite{hasan1992flow}). This compressed fluid is subsequently ejected, pushing shear layer outward and causing self-sustained flapping. The flapping motion occurs at a non-dimensional frequency of O(1) (\cite{spazzini2001unsteady}, \cite{wee2004self}). 

Liquid flows in the wake of the step can experience cavitation if the local pressure in the shear layer falls below the vapor pressure.  For turbulent shear flows, cavitation inception(defined as the first occurrence of observable cavitation) is known to first occur in streamwise braids as they are stretched between the spanwise vortices of the shear layer (\cite{katz1986cavitation} ; \cite{o1990experimental}). At this stage, cavitation bubbles are hypothesized to decouple the stretching and rotation rates of streamwise vortices (\cite{belahadji1995cavitation}) and known to not significantly alter bulk flow topology (\cite{iyer2002influence}).

As the extent of the cavitation increases, it can influence the vorticity in the shear layer, and gas can begin to fill the separated region of the flow.  For cavitating wakes, the presence of developed cavitation has a significant impact on large scale structure dynamics like vortex spacing, shedding frequency and convective speeds of spanwise structures (\cite{franc1983two}; \cite{young1966effects}; \cite{aeschlimann2011x}; \cite{wu2019bubbly}). 

 Classical re-entrant jet driven shedding and propagating bubbly shock-wave driven shedding are two known mechanisms that govern shedding dynamics in developed partial cavities (\cite{ganesh2016bubbly}; \cite{wu2019multimodal}). The re-entrant liquid flows are primarily driven by adverse pressure gradient at the cavity closure (\cite{callenaere2001cavitation}). The bubbly shockwaves on the other hand are known to occur due to the effect of compressibility of the vapor-liquid mixture that reduces the local speed of sound, thereby driving flow to go locally supersonic.   These features have also been observed in separated cavity flows formed behind obstacles (\cite{barbaca2019unsteady}). A proper expression for the speed of sound is necessary to accurately estimate the underlying dynamics of a rich bubbly shock, thereby explaining shock-wave induced instabilities that may effect cavity shedding. 

In the current study, we will examine the flow features of developed cavities formed in the wake of a backward facing step, and quantitatively understand the underlying cavity dynamics through high speed videography and time resolved X-ray densitometry. Our goal is to both gain a better understanding of the flow physics as well as to develop a data set that can be used for the validation of computational models. The manuscript is divided into five parts.  First, we present the experimental setup; we then define three flow regimes of interest (Type A, B and C) and describe their flow dynamics (cavity shedding frequency and pressure inside cavity). Finally we relate these dynamics to the underlying effect of flow compressibility.

\section{Experimental Setup}

\begin{figure}
	\centering
	\includegraphics[width=1\linewidth]{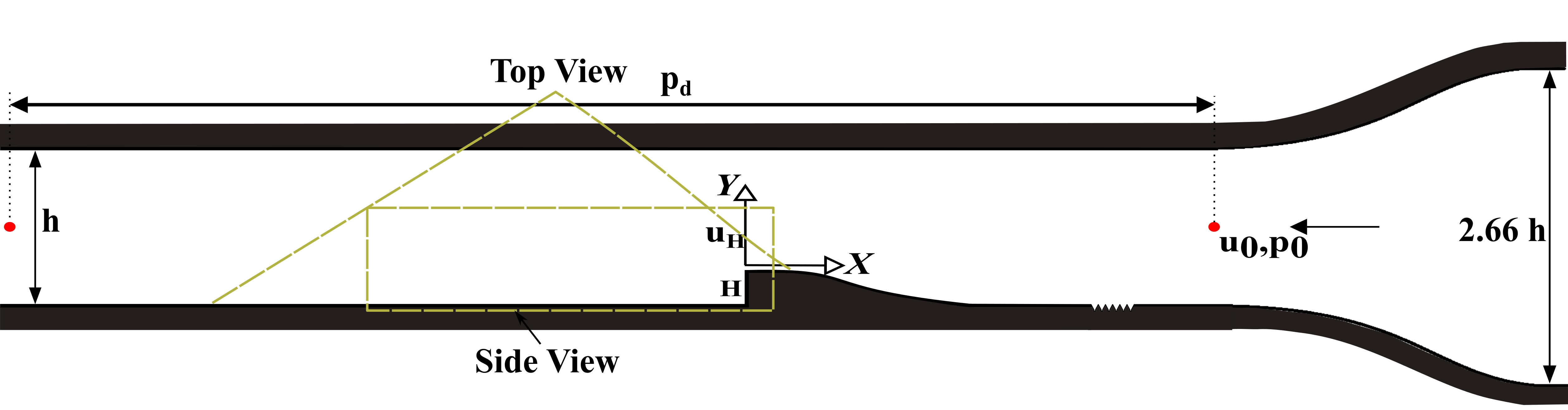}
	\caption{Schematic of the backward facing step mounted in the water tunnel. Top and side field of view configurations for the high speed videography are shown. For X-ray densitometry a 115 mm by 28 mm view, roughly similar to the highlighted side view is taken (see Figure \ref{fig:figure4p2typeatypebtypec})}. 
	\label{fig:figure2tunnelschematic}
\end{figure}

Experiments were carried out in the Michigan 9" recirculating cavitation channel. The upper leg of the tunnel has 6:1 contraction leading to 228.6 mm diameter section that leads to 209.6 mm by 209.6 mm square section with filleted corners. To minimize the baseline attenuation in X-ray densitometry for non-cavitating flow, the channel is further reduced to 78.4 mm by 78.4 mm test section. A fifth order polynomial profile (f(x)) leads to the flat ramp of the backward facing step. Step height (H) was chosen to match the blockage used by researchers at Johns Hopkins(\cite{agarwal2018cavitating}). The test section profile is shown in Figure \ref{fig:figure2tunnelschematic}. 

 The inlet velocity ($u_0$) was measured by measuring the pressure drop across the 6:1 contraction using Setra 230, 0-68 kPa, differential pressure transducer with an accuracy of 0.25\% of full scale (FS). From this measured $u_0$ velocity at step($u_{H}$) is determined using mass conservation($h*h*u_0$ = $(h-H)*h*u_H$) For current dataset $u_{H}$ of 8, 10 and 12 m/s are maintained for $Re_H$ = $8.3 \cdot{10^{4}}$, $10.6 \cdot{10^{4}}$ and $12.7 \cdot{10^{4}}$  respectively to operate the water tunnel within practical inlet pressures ($p_0$). Inlet pressure ($p_0$) was varied from inception to before super-cavitation for a $\sigma_0$ range of 0.5 to 1.0. $p_0$ was measured using Omega PX409-030A5V, 0-206 kPa static pressure transducer (0.08 \% FS accuracy). Pressure $p_1$ was measured using differential pressure transducer Omega PX2300, 0-34 kPa (0.25 \% FS accuracy) between pressure taps at $p_0$ and $p_1$. Omega PX409-005AI-EH, 0-34 kPa (0.08 \% FS accuracy) static pressure transducer measures pressure at X = 1.5H ($p_2$); while Omega PX203, 0-206 kPa (0.25 \% FS accuracy) static pressure transducer measures pressure at X = 8H ($p_3$). Dynamic pressure transducer PCB 113M231, 7.04 $\mu A$/kPa sensitivity is used to measure dynamic pressure \textit{dp}. The signal is acquired at a sampling frequency of 50 kHz for 3 seconds. Finally the overall pressure drop across the test section is measured using Omega RX2300, 0-34 kPa (0.25 \% FS accuracy). All pressure taps on optically accessible acrylic windows have ~1.6 mm diameter. All the static and dynamic pressure measurement locations are summarized in Table \ref{tab:my-table}. The dissolved oxygen was maintained below 20 \% of saturation for all experiments.

2-D planar PIV was done to characterize the single phase flow of the setup by determining the boundary layer thickness at the step and reattachment length. High speed cinematography (HS) was performed using time synchronized Phantom 710 cameras at 5000 fps for 1.5 s with both top and east side view. Cinematographic X-ray densitometry system (\cite{ganesh2016bubbly}, \cite{makiharju2013time}) was used to make spanwise-averaged quantitative void fraction ($\alpha  $) measurements at 1000 fps for 0.787 s exposure time and 0.125 mm spatial resolution.

\begin{table}
\includegraphics[width=1\linewidth]{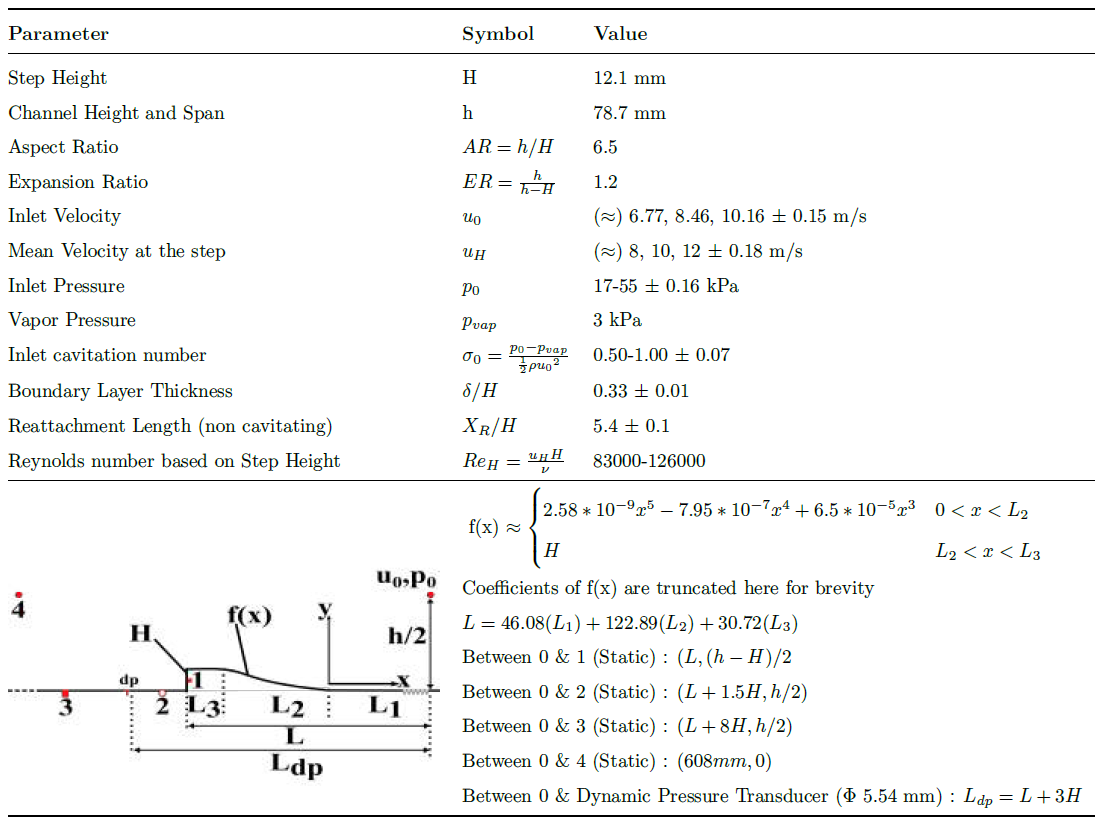}
\caption{Inlet parameters, transducer locations and geometry of the step.}
\label{tab:my-table} 
\end{table}

\section{Results}

\begin{figure}
	\centering
	\includegraphics[scale=0.9]{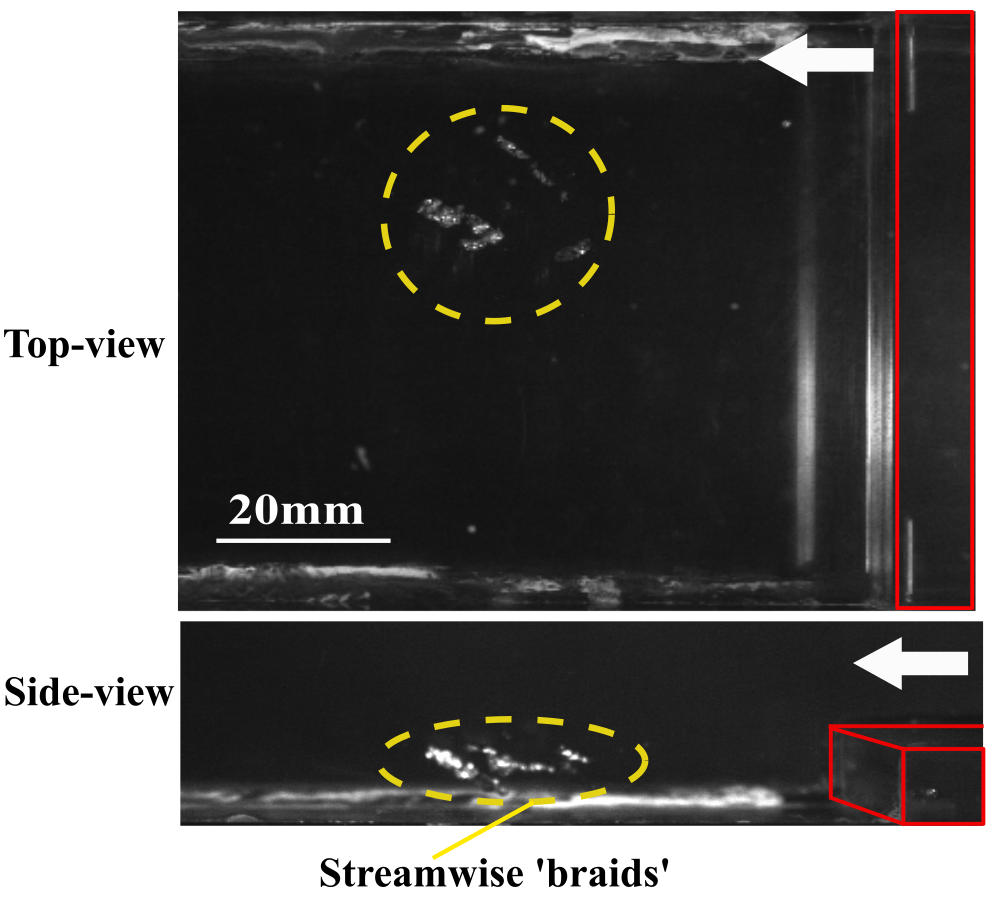}
	\caption{Cavitation inception in the streamwise direction is highlighted (streamwise 'braids' \cite{katz1986cavitation}). Intermittent cavitation occurs at $ \sigma_{0}=0.98$ for $u_{H}$ = 10 m/s. The red lines highlight the edge of the step. }
	\label{fig:figure2inception}
\end{figure}

\begin{figure}[ht]
	\centering
	  \includegraphics[scale=.9]{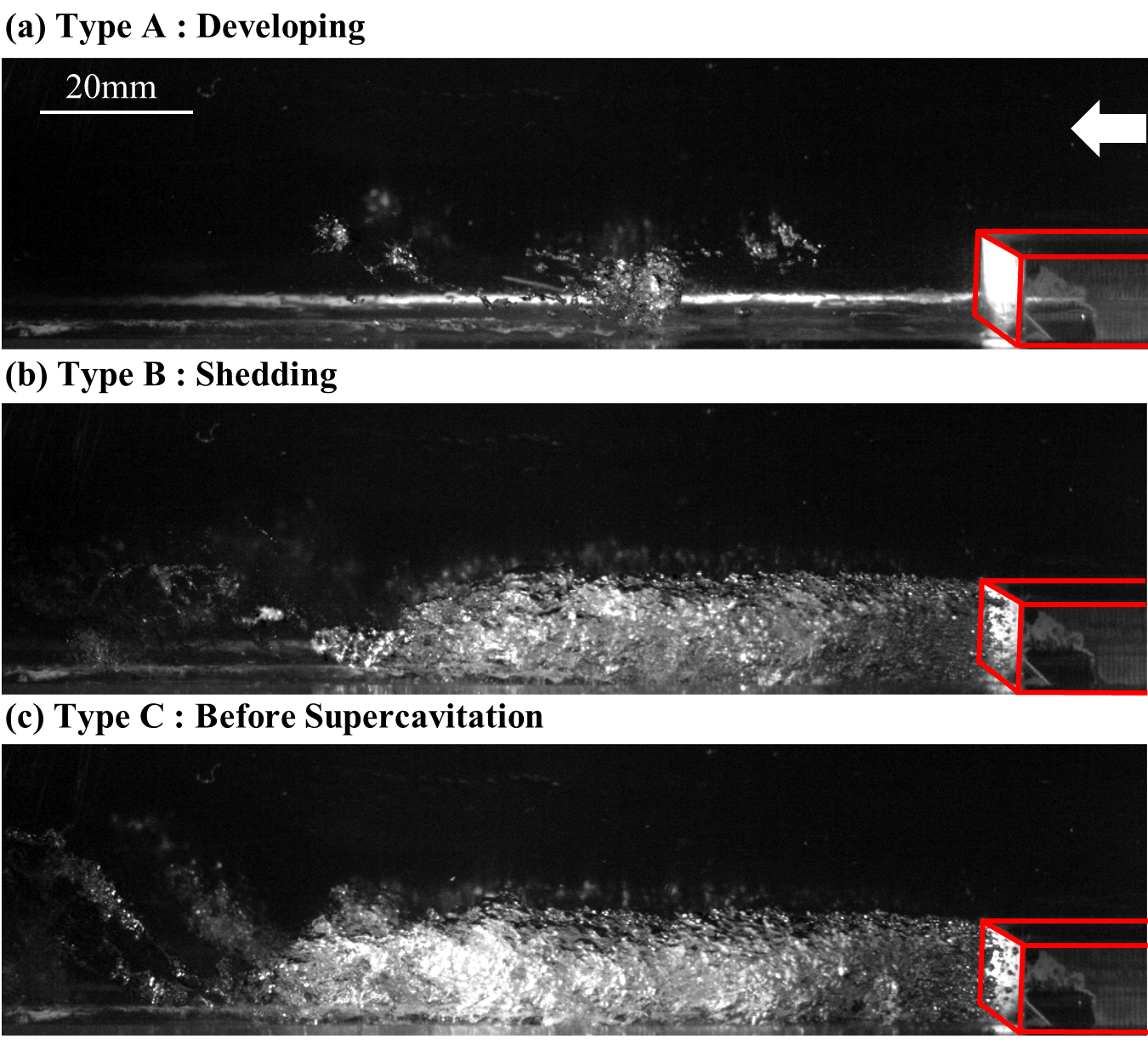}
	\caption{High speed snapshots of three different cavitation regimes (a, b, c for $\sigma_{0}$=0.78, 0.72, 0.60 respectively) at $u_{H}$ = 10 m/s. Type A comprises of cavitation in shear layer mostly near reattachment region(away from separation point). Types B and C shedding regimes are of primary interest in this study.}
	\label{fig:figure4p1typeatypebtypec}
\end{figure}

The experiments were conducted by keeping the velocity at step $ u_H $, fixed at 8, 10 or 12 m/s  and reducing inlet pressure ($ p_0 $) from inception (near $ \sigma_{0} = 1 $)  to near super-cavitation (near $~\sigma_{0} = 0.6 $). As seen in previous literature (\cite{o1990experimental}), inception first occurs in streamwise braids, away from the separation point. The inception condition is shown in Figure  \ref{fig:figure2inception}. On reducing $p_0$ further, greater extent of cavitation occurs primarily between X = -3H to -5H, classified as Type A or ‘developing cavitation’ phase of the flow. Developing cavitation is characterized by the presence of vapor bubbles in spanwise vortices of the shear layer. A high-speed video snapshot of developing cavitation is depicted in Figure  \ref{fig:figure4p1typeatypebtypec}(a).

\begin{figure}[ht]
	\centering
    \includegraphics[scale=0.9]{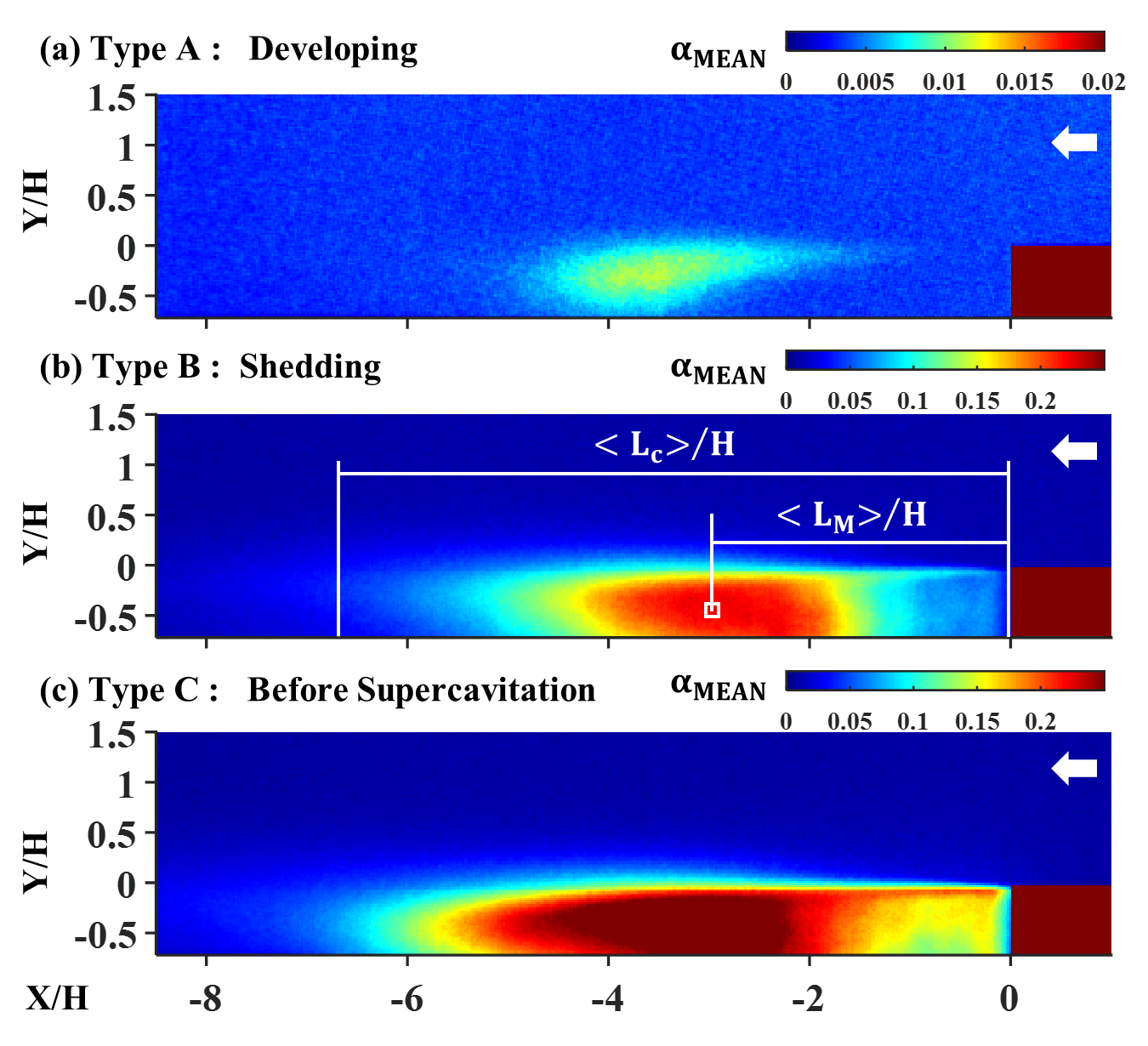}
	\caption{The mean void fraction fields (snapshots) of the three cavitation regimes (a, b, c for $\sigma_{0}$=0.78, 0.72, 0.60 respectively) at $u_{H}$ = 10 m/s. The mean length of cavity $<L_{C}> /H$ and distance to maximum void fraction value (in the mean field, $ <L_{M}> /H$ are shown. The actual densitometry data captures entire 0-100 \% void fraction scale, but color scale artificially limited to 0-2\% for Type A and 0-25\% for the rest. }
	\label{fig:figure4p2typeatypebtypec}
\end{figure}

With a reduction in inlet pressure $p_0$, unsteady cavitation designated as Type B is observed. Type B self-sustained shedding cavitation is the regime where periodic oscillation of the cavity length occurs at comparatively higher frequency (Typical cavity fluctuations for Type A occur below 10Hz). As $p_0$ is dropped even further, Type C cavitation is observed. This is the stage right before super-cavitation where shedding still occurs albeit at a much lower rate and cavity is filled with vapor. The instantaneous cavity lengths for Type C start to extend beyond the field of view of the X-ray densitometry system (roughly the side view shown in Figure \ref{fig:figure2tunnelschematic}). The three phases of cavitation are shown in Figure  \ref{fig:figure4p1typeatypebtypec} and Figure \ref{fig:figure4p2typeatypebtypec}.

For inception and developing cavitation regimes the measured mean void fraction values are less than 10\%. The X-ray densitometry system yields ~2\% uncertainty in the reported void fraction values. This is the main reason why higher void fraction Type-B and Type-C regimes are of primary interest in the current study (acceptable signal to noise ratio (SNR)).

\subsection{Average Cavity Topology}

We use the mean void fraction flow fields to determine the downstream location where void fraction values ($\alpha_{MEAN}$) are 0.05 in the region of cavity closure, defining the mean cavity length $<L_{C}>$. The mean void fraction field is estimated as :

\begin{equation}
    \alpha_{MEAN}  = \ \sum\limits_{N=0}^{786} \dfrac{\alpha(x,y,N)}{787}
\end{equation}

\begin{figure}
	\centering
	\includegraphics[scale=0.95]{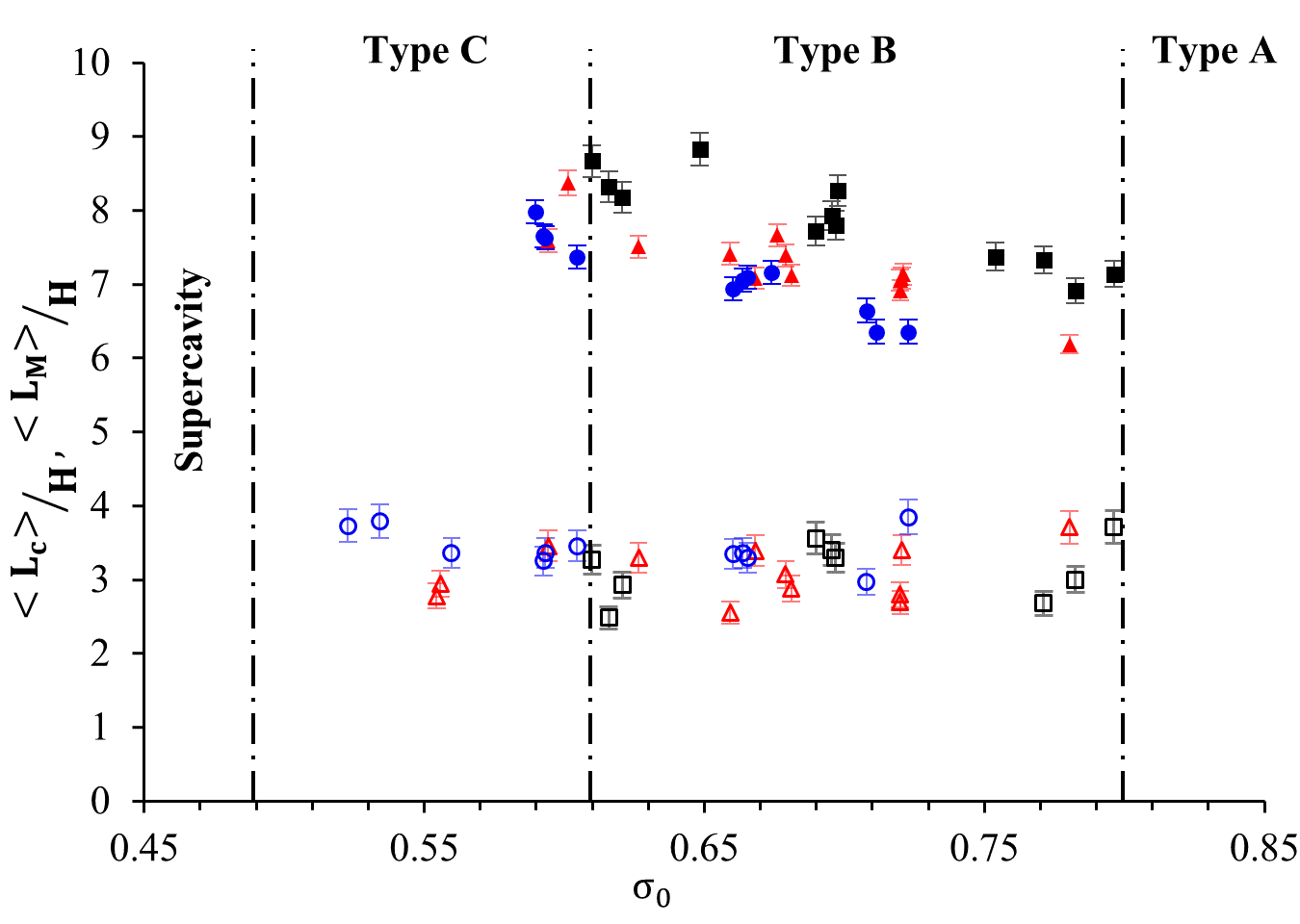}
\caption{Mean cavity length($<L_c>$) and location of maximum void fraction($<L_M>$) for $u_{H} = 8 m/s (\blacksquare, \square)$, $10 m/s (\color{red}\blacktriangle, \triangle)$, $12 m/s (\color{blue}\circ, \bullet)$ for different $\sigma_{0}$. Closed symbols are for $<L_c>$ and open symbols for $<L_M>$.}
	\label{fig:figure5cavitylength}
\end{figure}

The distance to the maximum void fraction ($max(\alpha_{MEAN}) $) in the field is recorded as $<L_{M}>$. Both these distances are measured from the separation point on the step, as shown in Figure  \ref{fig:figure4p2typeatypebtypec}(b) and plotted in Figure  \ref{fig:figure5cavitylength}.

 It is observed that higher the Re, for a given $ \sigma_{0} $, the lower the mean cavity length.  Conversely,  $ L_{M} $ lies between 3 to 4H, regardless of the Re. This can be juxtaposed with the shear stress distribution reported in the previous literature (\cite{jovic1995reynolds}). The shear stress maximizes roughly at 0.67 reattachment length, which is ~3.7H for their specific single phase flow configuration.  

\subsection{Typical Shedding Cycle}

\begin{figure}
	\centering
	\vspace*{-0.75in}
	\includegraphics[scale=0.85]{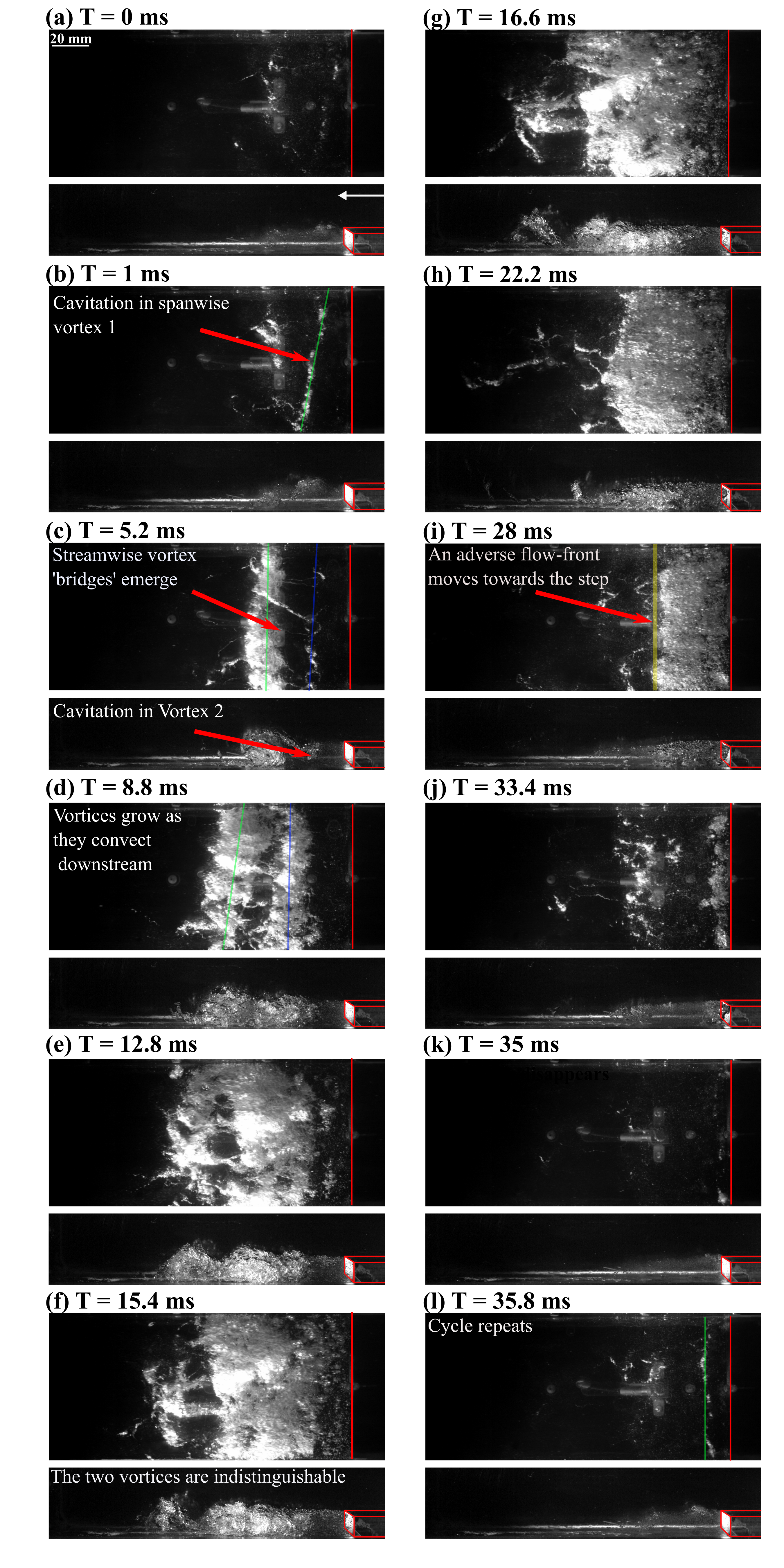}
	\vspace*{-0.2in}
	\caption{The timeseries of high speed snapshots of a periodic Type B shedding cycle. Case at $ \sigma_{0} $=0.72 for $ u_{H} $ = 10 m/s. Cycle starts with visible cavitation in spanwise vortex(b). The recirculation bubble and shear layer continue to fill with vapor(e)-(g), until an adverse flow front(i) causes the cavity to disappear(shed cavity in (k)). }
	\label{fig:figure6snip}
\end{figure}

To better identify specific flow features during the shedding cycle observed in Type B and Type C cavities, high speed cinematographic frames of the flow are shown in Figure  \ref{fig:figure6snip}. The primary cavitating spanwise vortex (green) and secondary vortex (blue) are seen interacting in Figure  \ref{fig:figure6snip}(c). This interaction can be seen through streamwise vortex bridges formed between the two spanwise vortices. Both these vortices have the same sense of rotation(anticlockwise). During the filling period (Figure  \ref{fig:figure6snip}(f)-(g)) the side-view of the cavity appears quasi-steady. The ejected spanwise vortex at the end of a shedding cycle can be visualized in Figure  \ref{fig:figure6snip}(l).

\begin{figure}
	\centering
	\includegraphics[width=0.8\linewidth]{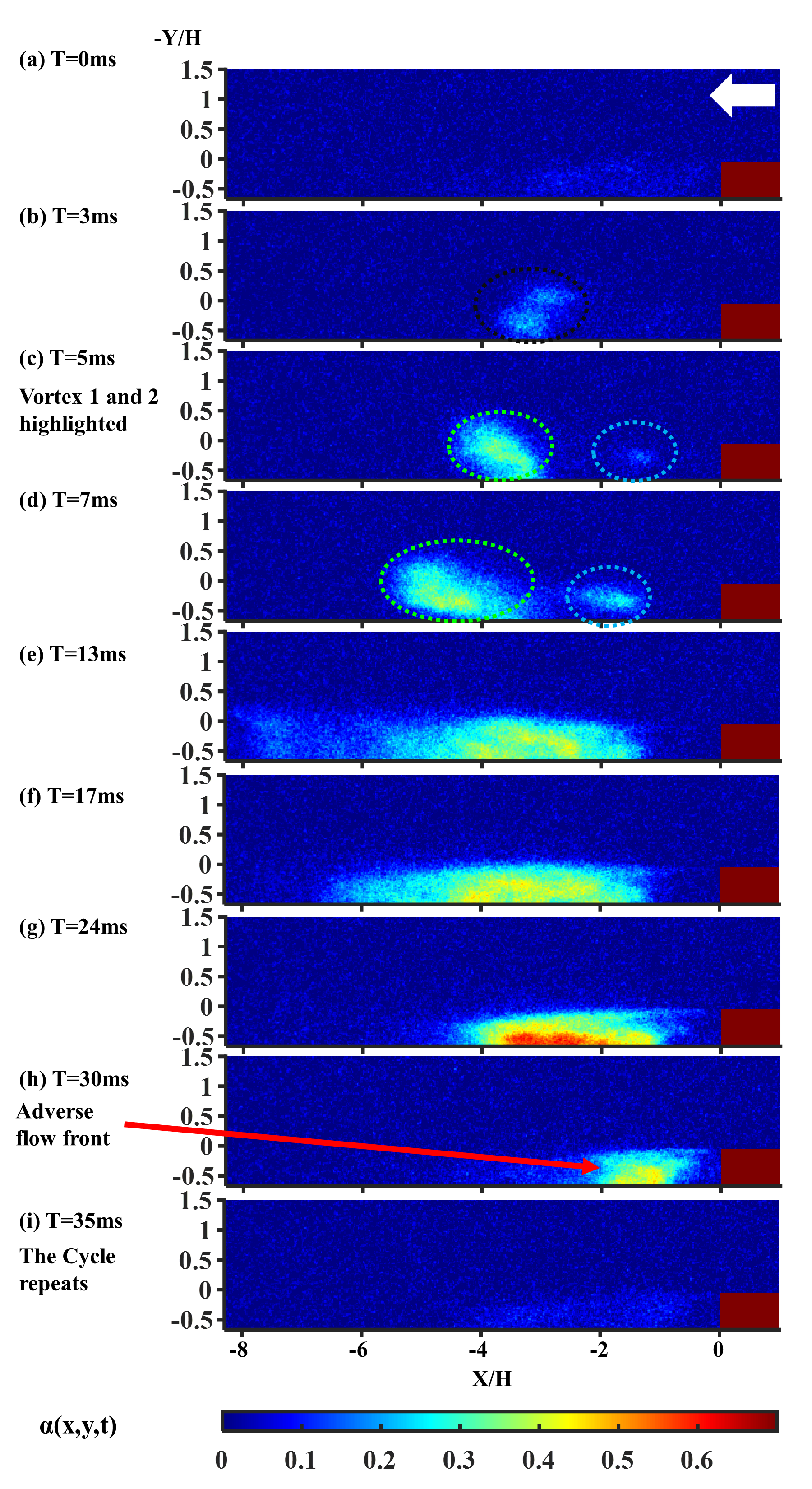}
	\caption{X-Ray timeseries of a typical Type B shedding cycle at $\sigma_{0}$=0.72 for $u_{H}$ = 10 m/s. The void fraction ($ \alpha(x,y,t) $) range is set to 0-70\%. The adverse flow front highlighted in (h) resembles the bubbly shock front  }
	\label{fig:figure7xrtimeseries}
\end{figure}

The X-Ray densitometry image frames of a typical Type-B shedding cycle are shown in Figure  \ref{fig:figure7xrtimeseries}. The cycle begins with cavitation in the core of a spanwise vortex (vortex 1), highlighted in green between X = -0.5H to X = -2H (Figure  \ref{fig:figure7xrtimeseries}(a)). As this structure convects downstream, it grows in size and interacts with another vortex (vortex 2, highlighted in blue). Subsequently the vortex pair becomes indistinguishable and a filled cavity starts to emerge between X = -2H to X = -6H seen in Figure  \ref{fig:figure7xrtimeseries}(d)-(f). The seemingly quasi-steady snapshots shown in Figure  \ref{fig:figure6snip}(f)-(g) can be compared to Figure  \ref{fig:figure7xrtimeseries}(e)-(f) to indicate additional flow features captured through densitometry.  The cavity consistently grows from the aft, i.e. it fills from roughly around the reattachment zone until a filled cavity state is achieved.

Once a fully filled cavity is formed and reaches a high instantaneous void fraction value($\alpha=0.5-0.8$), a void fraction front begins to move upstream, resulting in a cavity shedding seen in Figure  \ref{fig:figure7xrtimeseries}(g)-(h).  As this flow front reaches the step, the cavity disappears and a cavitating spanwise vortex (green) is ejected into the oncoming flow. The shedding cycle then starts all over again. 

\begin{figure}
	\centering
	\includegraphics[width=0.8\linewidth]{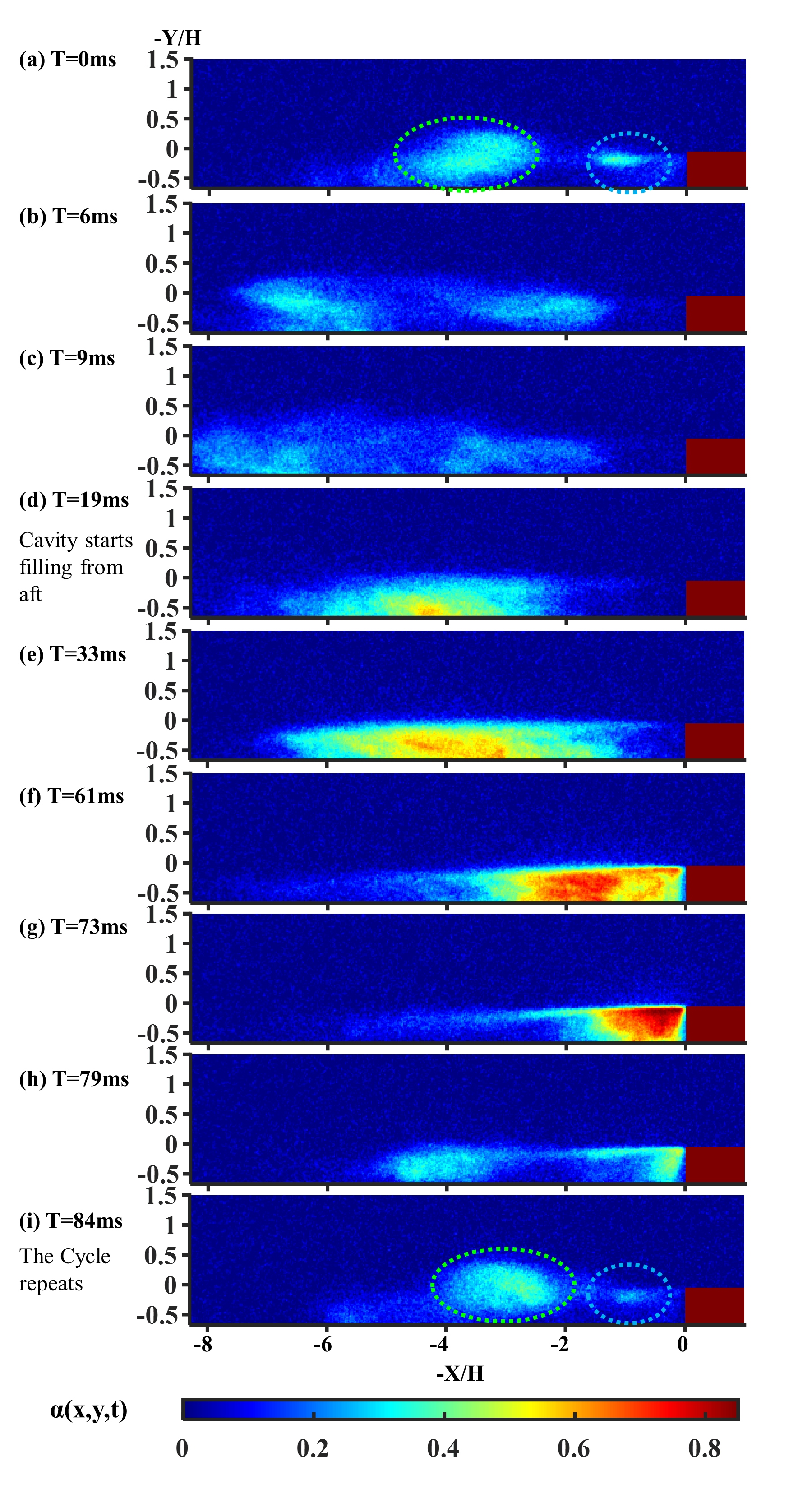}
	\caption{X-Ray timeseries of a typical Type C shedding cycle at $\sigma_{0}$=0.60 for $u_{H}$ = 10 m/s . The void fraction ($ \alpha(x,y,t) $) range is set to 0-85\%. Compared to Type B, Type C cavities are longer, have higher vapor content and collapse at a slower rate. High speed snapshots of Type C appear ~quasi steady, unlike the X-Ray timeseries which clearly show shedding.  }
	\label{fig:figure7xrtimeseries_typeC}
\end{figure}

The Type C shedding case is depicted in Figure \ref{fig:figure7xrtimeseries_typeC}. Most features of the cavitating cycle are similar to those of Type B albeit at a much slower rate that makes it ~quasi-steady in direct high speed videos. The lower cavitation number leads to fuller/longer cavity with a consistently filled shear layer.   

\subsection{Void Fraction}
\begin{figure}
	\centering
	\includegraphics[width=0.7\linewidth]{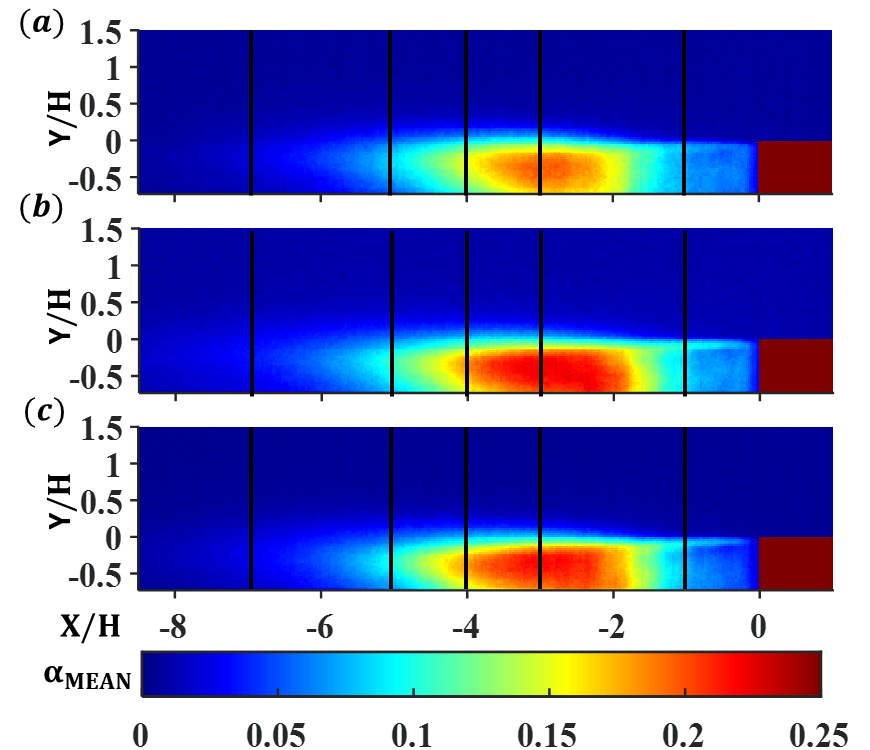}
    	\caption{Mean void fraction contours at $ u_{H} $ = (a) 8 m/s, (b) 10 m/s and (c) 12 m/s at $\sigma_{0} = 0.72$. 5 slices at $X/H=$ -1, -3, -4, -5, -7 are taken to compare void fraction values across the three velocities shown in Figures \ref{fig:figure9meanvf},\ref{fig:figure10meanvftypec}, \ref{fig:figure11rmsdvftypebandc}.}
	\label{fig:figure8slices}
\end{figure}

Based upon local pressure distribution and flow dynamics, void fraction values can vary significantly. The mean void fraction profiles for the onset of Type B cavitation at different Re are shown in Figure  \ref{fig:figure8slices}. It can be seen that the mean void fraction profiles for the onset of shedding are almost identical across different Re.

\begin{figure}
	\centering
	\includegraphics[width=0.7\linewidth]{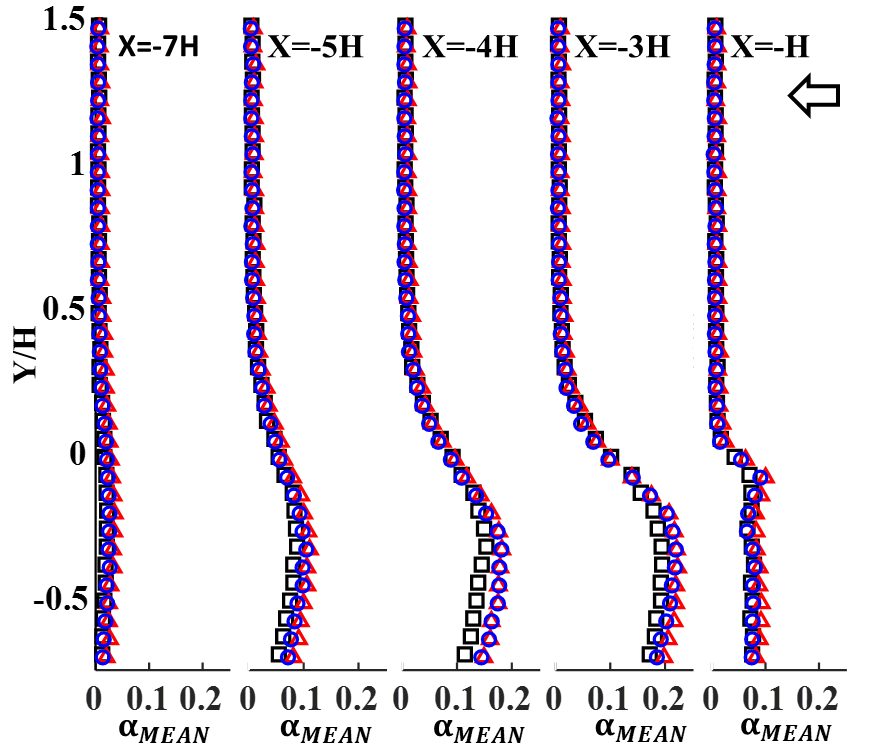}
	\caption{Mean void fraction values(2\% uncertainty) along slices Y/H = -1 to 1.5; $u_{H} = 8 m/s (\square)$, 10 m/s ($\color{red}\triangle $) and $12 m/s (\color{blue}\circ $) at $\sigma_{0} = 0.72$ (Type B). The mean void fraction values are comparable across three Re. }
	\label{fig:figure9meanvf}
\end{figure}

\begin{figure}
	\centering
	\includegraphics[width=0.7\linewidth]{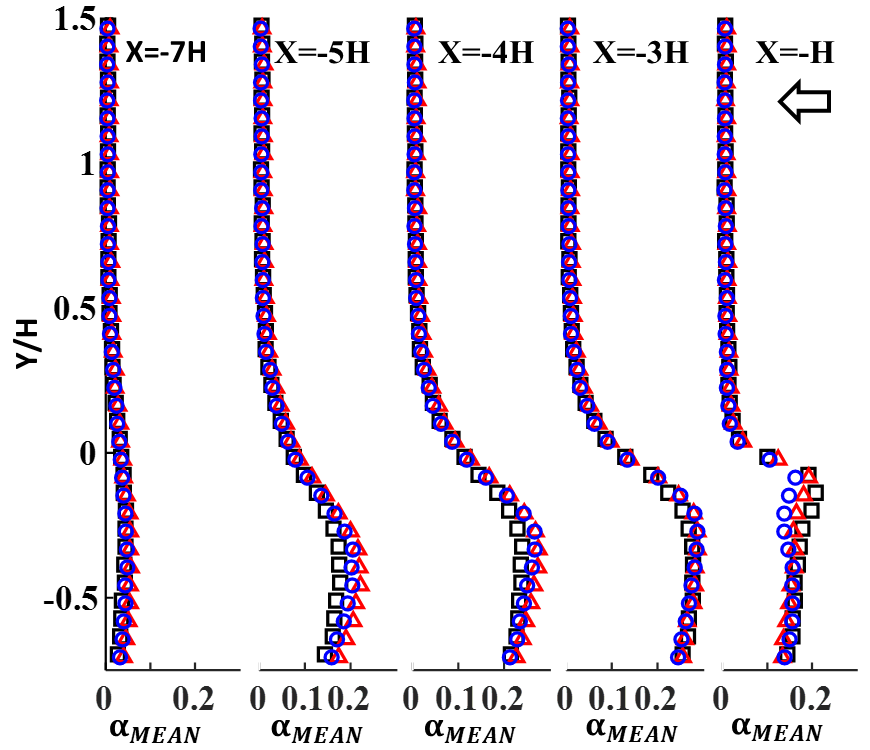}
	\caption{Mean void fraction values for Type C; $u_{H}$ = 8 m/s ($\square$), 10 m/s ( $\color{red}\triangle $ ) and 12 m/s ($\color{blue}\circ $) at $\sigma_{0} = 0.60$. No significant change in void fraction values due to Re is seen. }
	\label{fig:figure10meanvftypec}
\end{figure}

\begin{figure}
	\centering
	\includegraphics[width=0.7\linewidth]{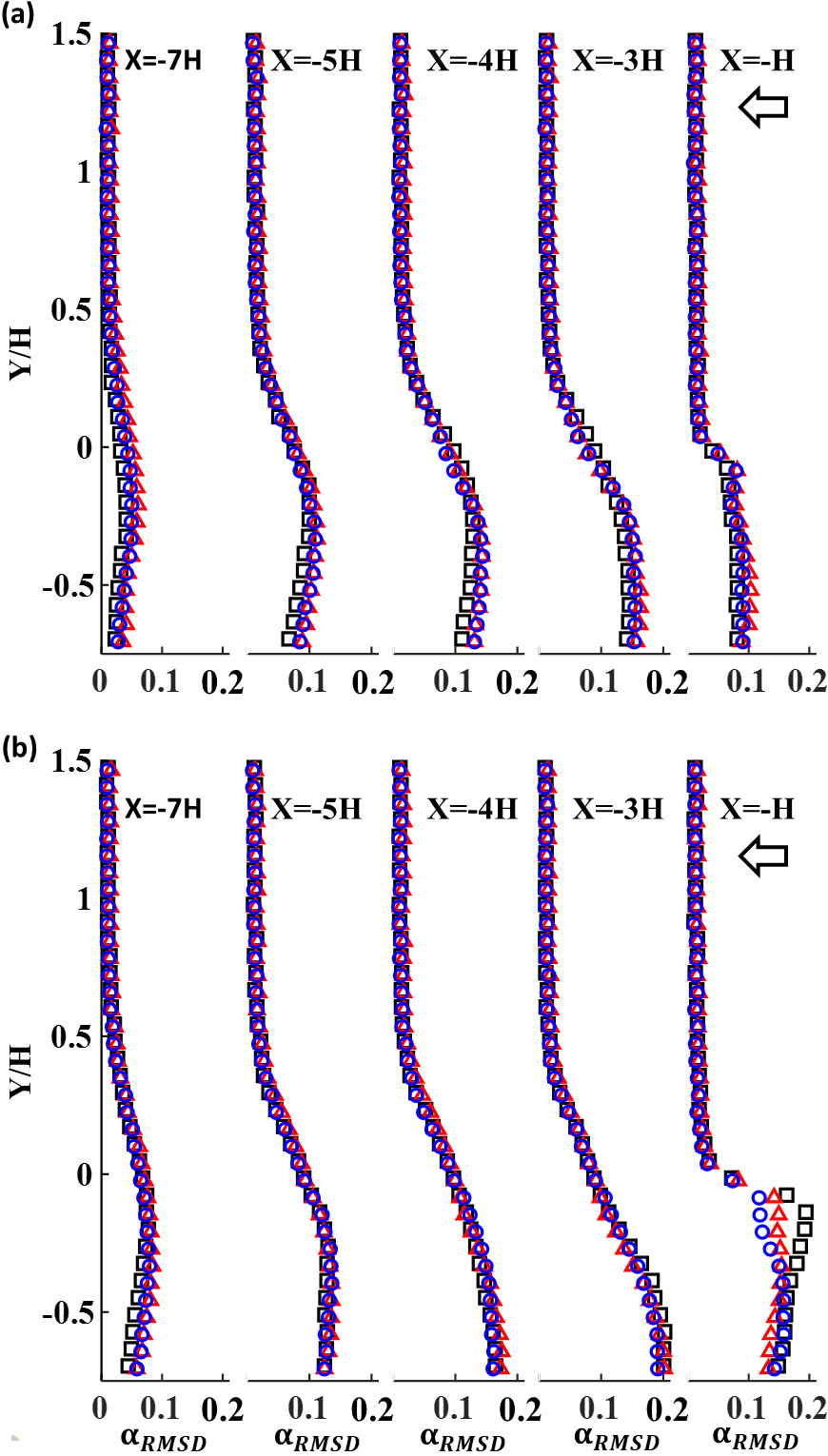}
	\caption{RMSD values for the three Re; $u_{H} = 8 m/s (\square)$, 10 m/s ( $\color{red}\triangle $ ) and $12 m/s (\color{blue}\circ $);(a) Type B;(b) Type C. RMSD values for Type B are comparable, meanwhile for Type C, RMSD of cavity void fraction is sensitive to Re. This can be partially attributed to different mean pressures at these Re that affect local pressure fluctuations.}
	\label{fig:figure11rmsdvftypebandc}
\end{figure}

X-slices of the mean void fraction field for Type B shedding regime at X = -H, -3H, -4H, -5H and -7H are shown in Figure  \ref{fig:figure9meanvf}. The mean void fraction values are comparable across different Re. X-slices slices for the Type C shedding are shown in Figure  \ref{fig:figure10meanvftypec}. The mean void fraction values, now higher in magnitude are still similar across Re. Taking similar slices on the Root Mean Standard Deviation (RMSD) void fraction field yields comparable RMSD values across the three Re for Type B regime in Figure  \ref{fig:figure11rmsdvftypebandc}(a). The RMSD field is generated as follows :
\begin{equation}
    \alpha_{RMSD} \  = \ \sqrt{ \dfrac{{\sum\limits_{N=0}^{786}(\alpha(x,y,N)-\alpha_{MEAN})}^2}{787}}
\end{equation}
Comparatively larger RMSD values are seen in Type C shedding regime (Figure  \ref{fig:figure11rmsdvftypebandc}(b)). With decreasing cavitation number, void fraction fluctuations increase. As mentioned earlier, the reported void-fraction values have ~2\% uncertainty in measurements.

\subsection{Cavity Shedding Frequency}

The cavity shedding cycle highlighted in Figure  \ref{fig:figure6snip} and Figure  \ref{fig:figure7xrtimeseries} occurs periodically, and as the flow transitions from Type A all the way to Type C, the shedding frequency, f,  also changes. The Strouhal number (St) used to non-dimensionalize the shedding frequency is given by $St = \frac{f \cdot{u_{0}}}{H}  $

\begin{figure}
	\centering
	\includegraphics[width=1.02\linewidth]{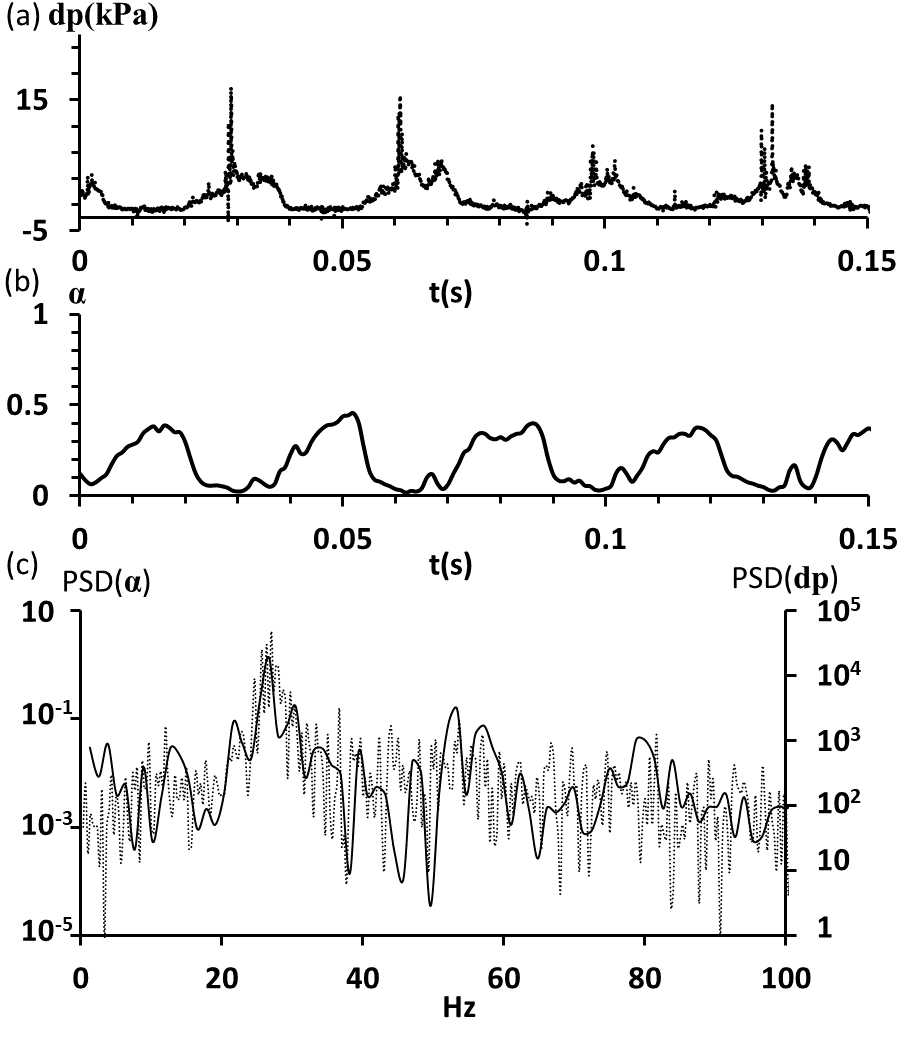}
	\caption{Dynamic pressure signal(a,$\cdot\cdot\cdot$ ) synchronised with time varying void fraction signal(b,$-$) and corresponding power spectral density(c) for a Type B cavitating case at $\sigma_0=$ 0.73 at $ u_H $ = 10 m/s. The highest energy peaks are close.}
	\label{fig:figure12sheddingrawsignalstypeb}
\end{figure}

\begin{figure}
	\centering
	\includegraphics[width=1.02\linewidth]{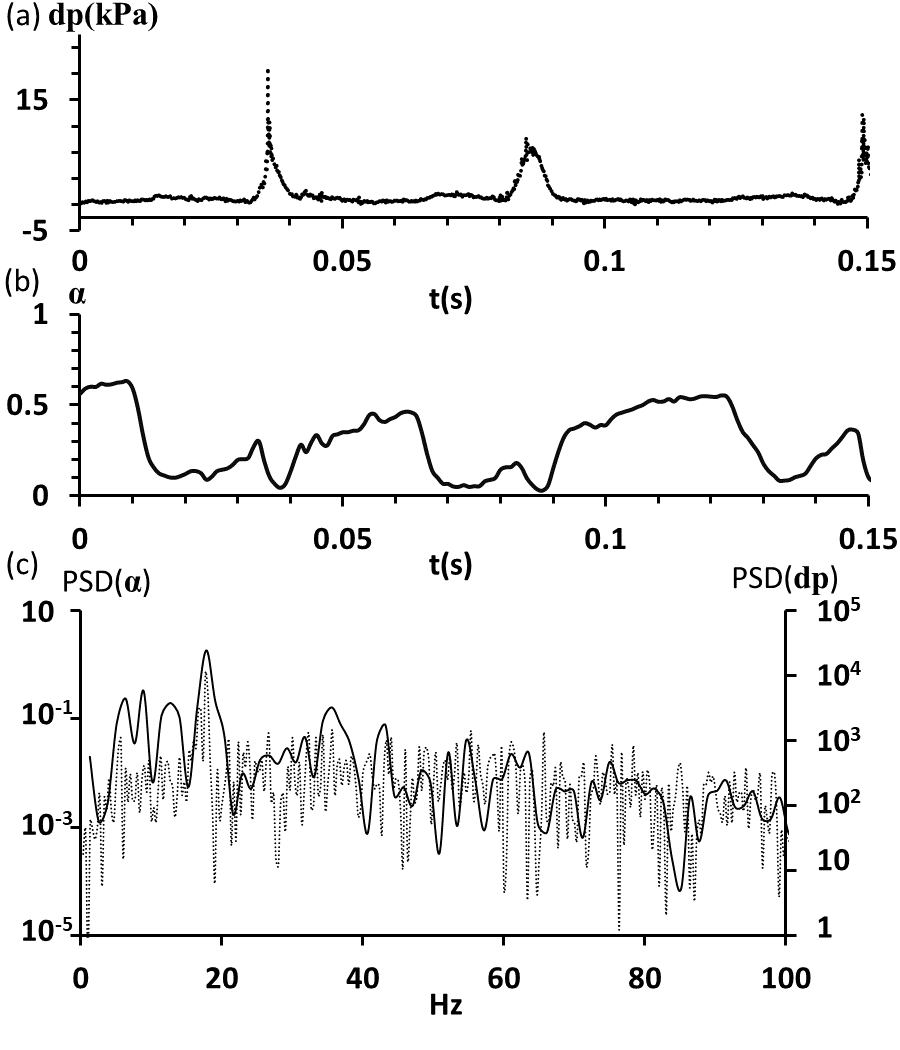}
	\caption{Dynamic pressure signal(a,$\cdot\cdot\cdot$ ) synchronised with time varying void fraction signal(b,$-$) and corresponding power spectral density for a Type C cavitating case at $\sigma_0=$ 0.59 at $ u_H $ = 10 m/s. Compared to Type B we can see a lower shedding frequency and lower pressures \textit{dp}.}
	\label{fig:figure12sheddingrawsignalstypec}
\end{figure}

\begin{figure}
	\centering
	\includegraphics[scale=0.9]{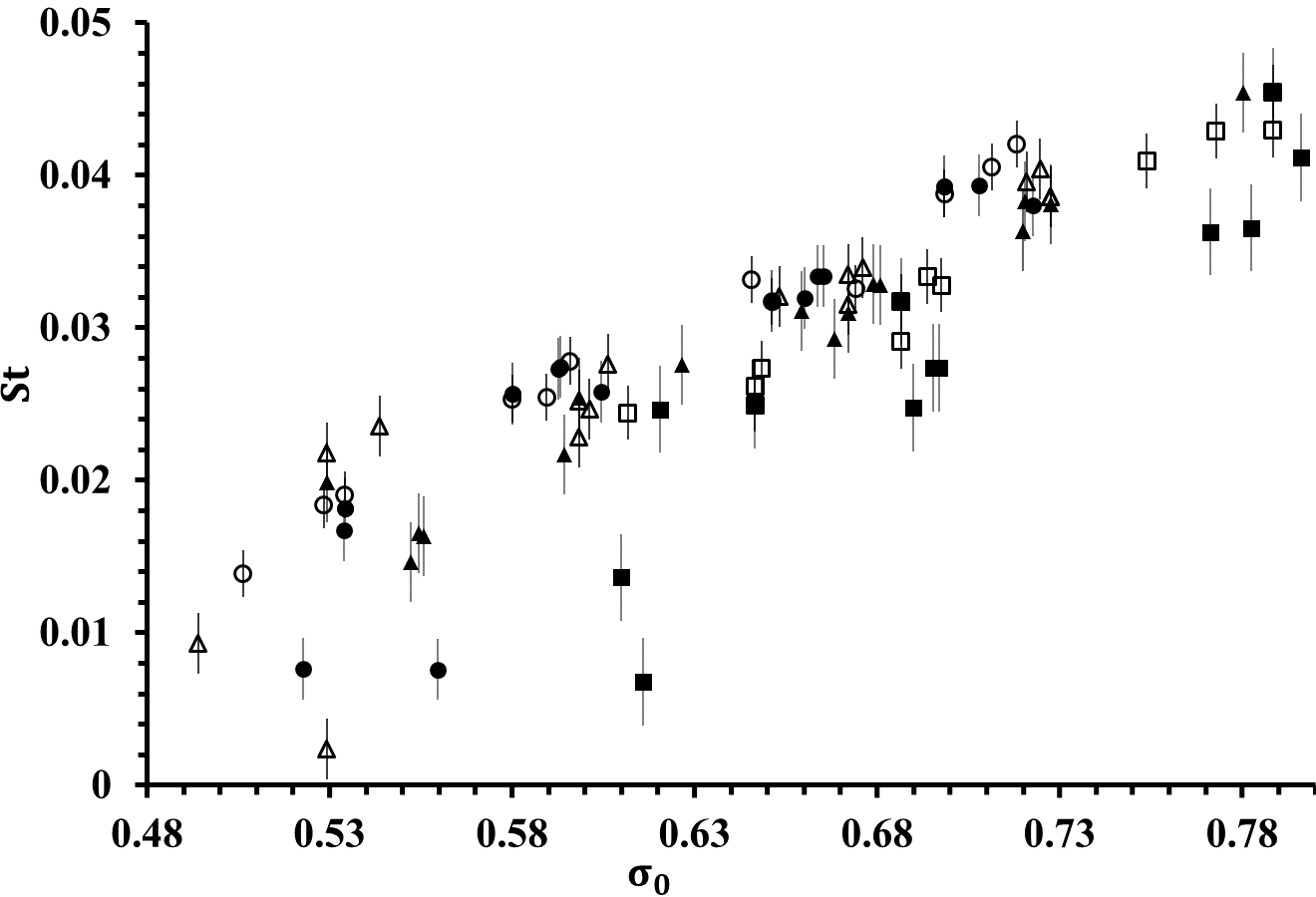}
	\caption{The change in strouhal number(shedding frequency) with decreasing $ \sigma_{0}$. $u_H =$ 8 m/s ($\square \blacksquare$), 10 m/s ( $\triangle \blacktriangle $) and $12 m/s$ ( $\circ \bullet $). The strouhal numbers obtained through the void fraction signals($\alpha$) are shown in filled symbols, while open symbols depict strouhal number from \textit{dp} signal. }
	\label{fig:figure13sheddingfrequency}
\end{figure}

The shedding frequency is determined from two time varying signals.  First, the spatially averaged local void fraction averaged over a 1.25mm by 0.875mm probe area located at (-5.5H, -0.5H). The shedding frequency observed through the void fraction signal does not change, for the probe located anywhere between X = -6.5H to -0.5H (Y= -H to 0).  Second, the shedding frequency is recorded with the dynamic pressure transducer (\textit{dp}). The dynamic pressure signal(\textit{dp}), the time varying void fraction signal and corresponding power spectral densities for a Type B and Type C case are shown in Figure  \ref{fig:figure12sheddingrawsignalstypeb} and Figure  \ref{fig:figure12sheddingrawsignalstypec} respectively. The primary shedding frequencies lie between 10-70 Hz. The resulting Strouhal number is defined with the strongest frequency peak, and it monotonically decreases with decreasing $\sigma_0$, shown in Figure  \ref{fig:figure13sheddingfrequency}. Note that St is not strongly influenced by changes in Re.

\subsection{Cavity Static Pressure}

The static wall pressure is measured at different streamwise locations beneath the cavity. The inlet pressure $p_0$ is used as the basis to find non-dimensional coefficient of pressure $Cp_i  = \frac{p_i  – p_0}{{\textstyle \frac{1}{2}} \cdot {\rho} \cdot{{u_{0}}^2}}$ where i=1, 2, 3 correspond to the pressure measured at the three static pressure taps beneath the cavity . 
\begin{figure}
	\centering
	\includegraphics[scale=1]{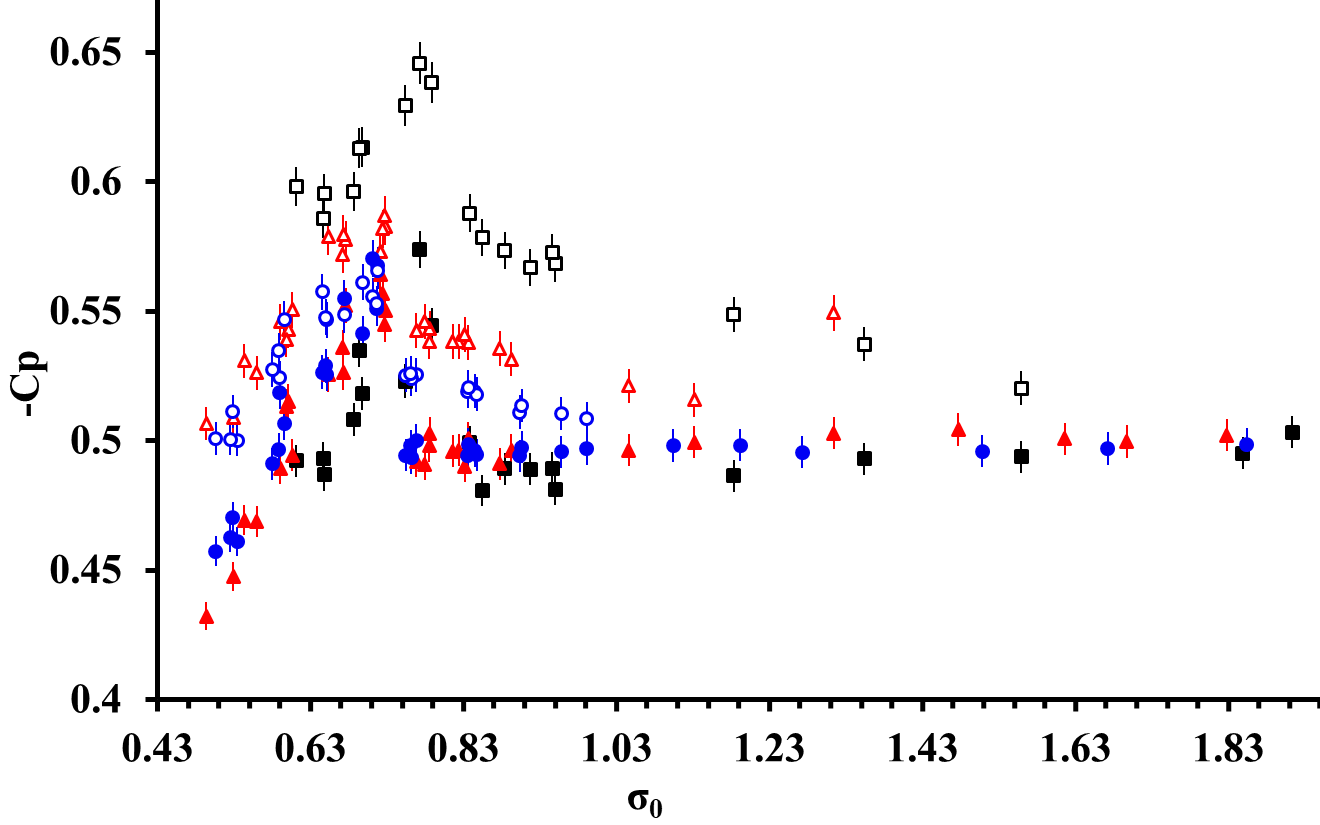}
	\caption{The values of $-Cp_1$ (filled, $ \blacksquare \color{red}\blacktriangle \color{blue}\bullet $) and $-Cp_2$ (open, $\square \color{red}\triangle \color{blue}\circ $ ) as $\sigma_0$ is reduced across three Re; $u_H =$ 8 m/s ($\square \blacksquare$), 10 m/s ( $\color{red}\triangle \blacktriangle $) and $12 m/s$ ( $\color{blue}\circ \bullet $). Magnitude of $Cp_2$ is higher than $Cp_1$ for all three Re indicating that behind the step, the pressure drops between pressure taps at location 1 and 2.}
  
 \label{fig:figure14cp1andcp2}
\end{figure}

\begin{figure}
	\centering
	\includegraphics[scale=1]{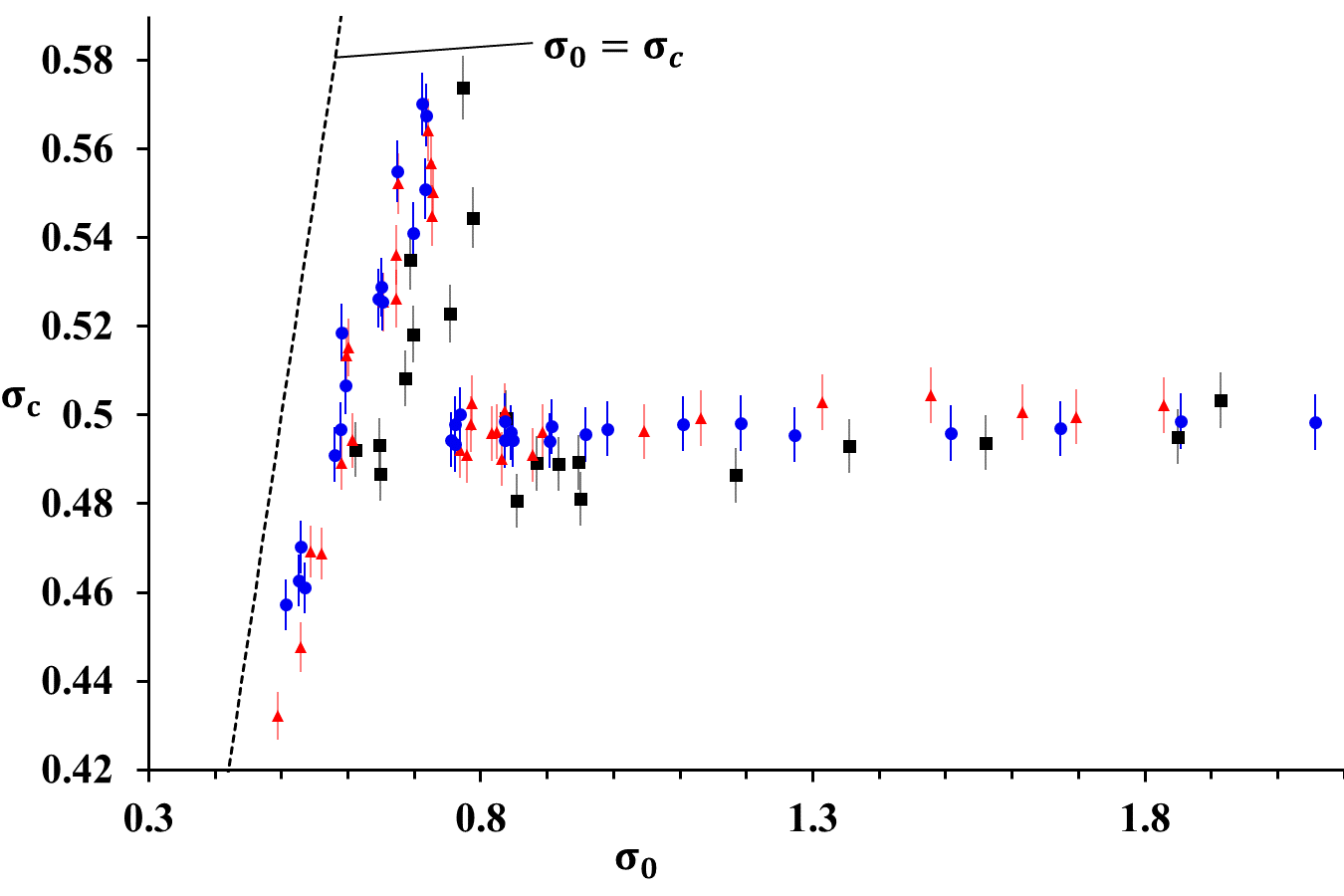}
	\caption{ $ \sigma_{c} $ with decreasing $ \sigma_{0} $ (Closer look at -$Cp_1$). $u_{H} = 8 m/s (\blacksquare$ ), $10 m/s (\color{red}\blacktriangle$), $12 m/s (\color{blue}\bullet$). At lower cavitation numbers, pressure inside the cavity approaches vapor pressure.}
	\label{fig:figure15sigmacvssigma0}
\end{figure}

The Cp values at different $\sigma_0 $ for $p_3$ and $p_2$ are shown in Figure  \ref{fig:figure14cp1andcp2}. $ -Cp_2 $ values are higher than $ -Cp_1 $ values across all three Re. Both $Cp_1$ and $Cp_2$ values peak around $ \sigma_{0} $ = 0.71-0.78, which corresponds to onset of Type B shedding. As cavity fills with vapour ($ \sigma_{0} $ is reduced), the $ -Cp $ value is seen to decrease. We can define a cavitation number based on cavity pressure as ($ \sigma_{c} $ ; where $ \sigma_{c} $ = $-Cp_1$. In Figure  \ref{fig:figure15sigmacvssigma0} we see that the cavity pressure approaches vapor pressure (--) as $ \sigma_{0} $ is reduced. This means that even during Type B and Type C shedding cycles, the cavity is not homogeneously filled with vapor, as also evidenced by the void fraction measurements.

\begin{figure}
	\centering
	\includegraphics[scale=1]{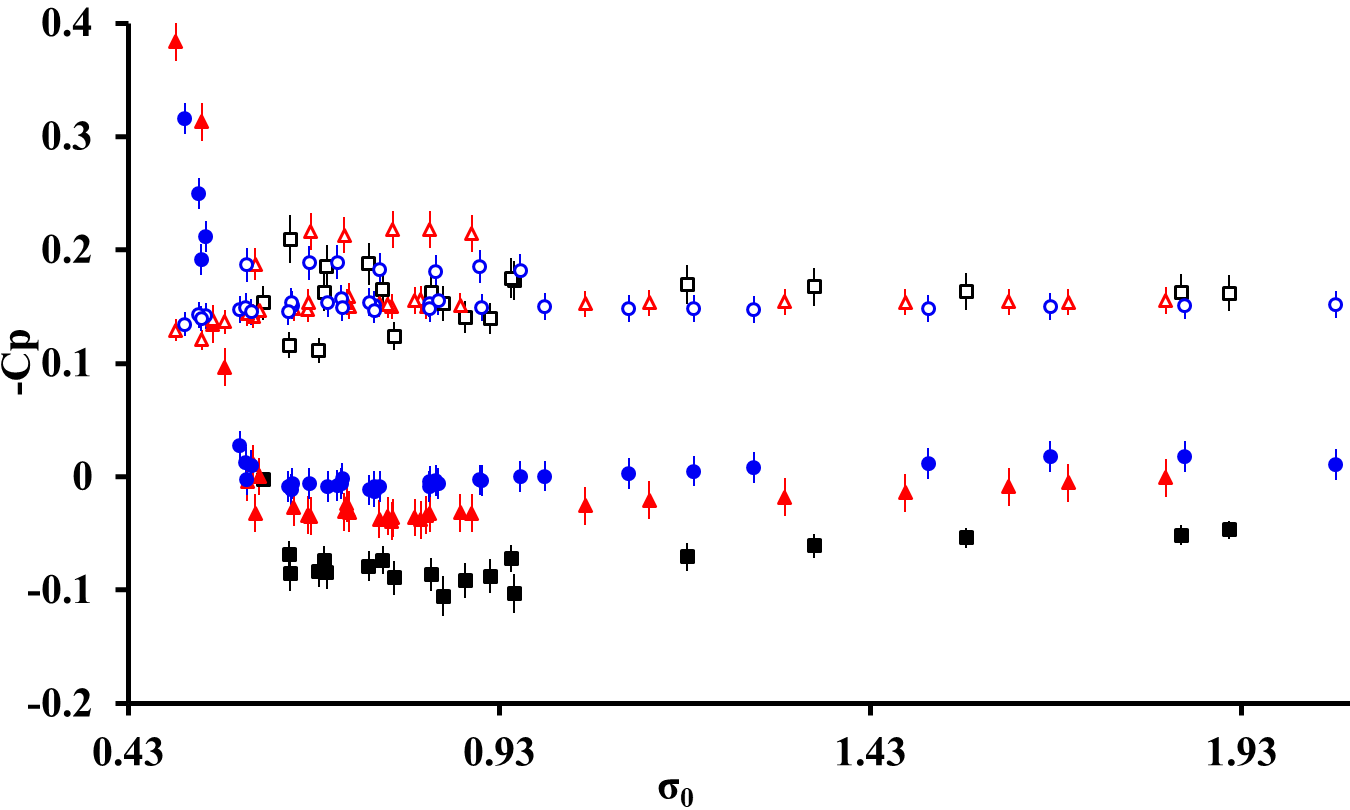}
	\caption{The values of $-Cp_3$ ($ \blacksquare \color{red}\blacktriangle \color{blue}\bullet $) and $-Cp_4$ ($\square \color{red}\triangle \color{blue}\circ $) as $\sigma_0 $ is dropped across three Re; $u_H = 8 m/s (\square \blacksquare)$, $10 m/s ( \color{red}\triangle \blacktriangle $) and $12 m/s ( \color{blue}\circ \bullet $ )}
	\label{fig:figure16cp3andcp4}
\end{figure}

-Cp for $p_3$ and $p_4$ are shown in Figure  \ref{fig:figure16cp3andcp4}. $Cp_4$ remains comparatively unchanged across different cavitation regimes. At high cavitation numbers $ Cp_3 $ is negative, indicative of higher pressure in the re-attachment region compared to the inlet pressure as suggested in previous literature(\cite{jovic1995reynolds}, \cite{ramamurthy1991some}). $-Cp_3 $ increases drastically as flow transitions from Type B to Type C. This is due to drop in $p_3$ as the cavity grows larger ($p_3$ approaches vapour pressure).

\subsection{Shockwave Properties}

 The flow front shown in Figure  \ref{fig:figure7xrtimeseries}(g) primarily drives the shedding(collapse) of the cavity, and this flow front is analogous to the propagating shockwaves in rich bubbly mixtures reported in previous literature (\cite{ganesh2016bubbly}). As $\sigma_0 $ is reduced (flow transitions from Type A to B) the vapor content in shear layer and recirculation bubble increases. The bulk modulus of the vapor liquid mixture decreases with decreasing $\sigma_0$ leading to higher compressibility in the shear layer and the recirculation bubble. Also with increasing vapor content the local speed of sound in the mixture decreases (\cite{brennen2005fundamentals}), and at optimal void fraction values the flow becomes locally supersonic. This makes the flow susceptible to formation of bubbly shock fronts that can govern overall the cavity shedding mechanism. 
 
\begin{figure}
	\centering
	\includegraphics[scale=1]{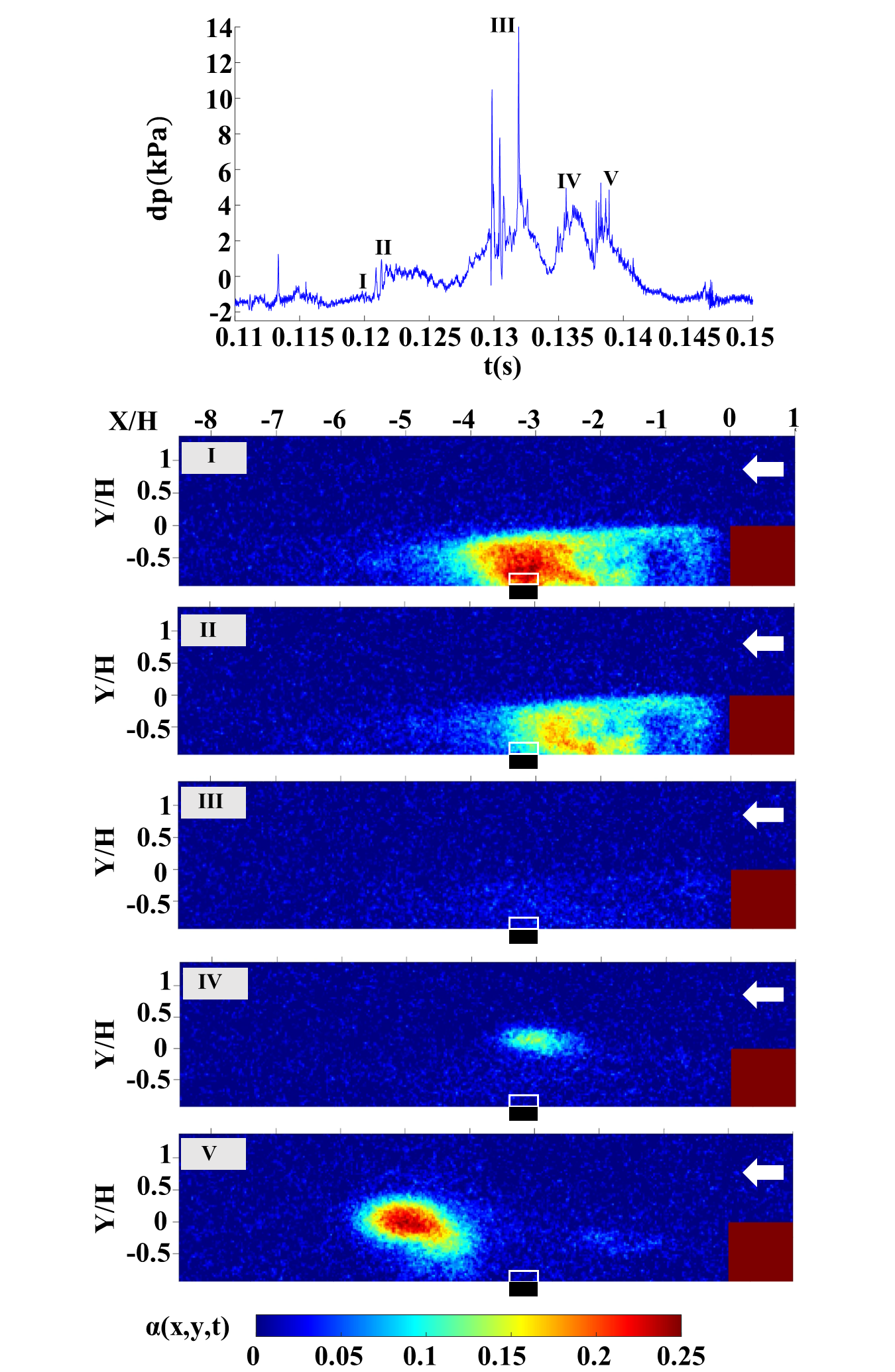}
	\caption{The differential pressure (\textit{dp}) and corresponding X-ray densitometry snapshots as the shock-front moves past the transducer and hits the step during a shedding cycle. $ \sigma_{0} $=0.72 for $ u_H $ = 10 m/s.}
	\label{fig:figure17propagatingshockfront}
\end{figure}

 As the cavitating spanwise vortices (highlighted in Figure \ref{fig:figure7xrtimeseries}(c),(d)) and subsequent bubbly shock fronts propagate over the differential pressure transducer (\textit{dp}) corresponding pressure rise can be recorded. This is shown in Figure  \ref{fig:figure17propagatingshockfront}, which highlights the different pressure rises over the dynamic pressure transducer (\textit{dp}) and the local void fraction probe.

Between stages I and II, a pressure rise is recorded, with the differential pressure \textit{dp} rising with the passage of the front.   When the flow front impinges on the step, another larger pressure pulse is created (III). The collapse of the shed vortices cause the remaining pressure spikes (IV and V). 

The speed at which this flow front propagates ($ U_{RH} $) can be predicted by the one dimensional Rankine-Hugoniot jump condition found by simple mass and momentum conservation across a 1-D shockwave in the shock-wave frame of reference:

\begin{equation}
U_{RH} =  \frac{d\rho}{\rho_L} \left[\frac{(1-\alpha_2)}{(1-\alpha_1)*(\alpha_1-\alpha_2)}\right] 
\label{eqn:shockspeed}
\end{equation}

where, $ \rho_L $=density of liquid(water) and $ \alpha_1$, $\alpha_2 $ are void fraction values pre and post shock. Since we have measured these values, we can compare the predicted and measured front propagation speed.  The $ \alpha_1$ and $\alpha_2 $ values are found by spatially averaging void fraction values in void fraction probe as the front convects over the dynamic pressure transducer, creating the corresponding change in pressure, \textit{dp}.

\begin{figure}
	\centering
	\includegraphics[scale=1.3]{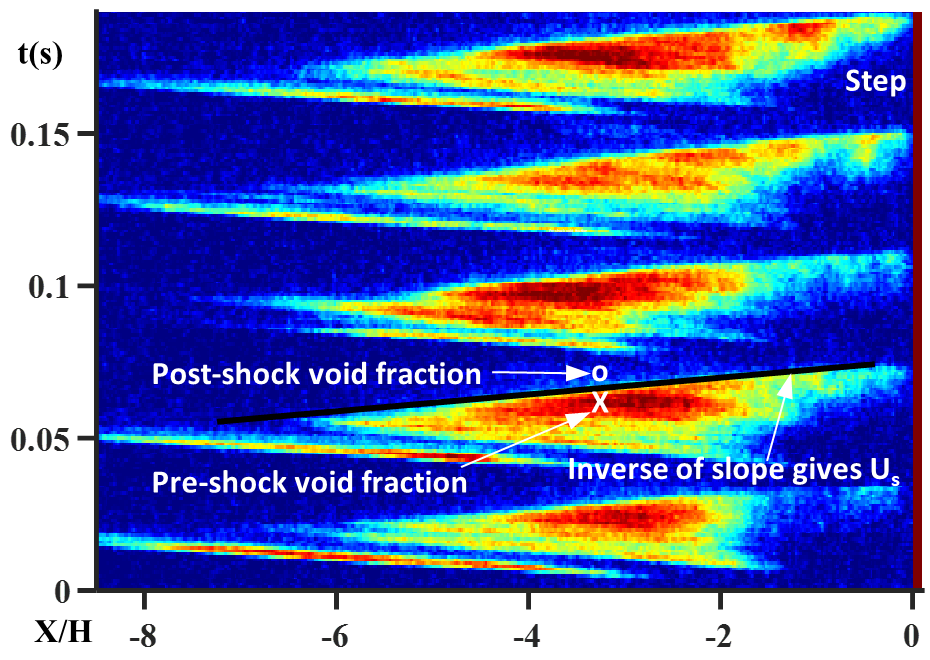}
	\caption{X-T diagram at Y/H = -0.25; $ \sigma_{0} $=0.72 for $ u_H $ = 10 m/s. The shock propagation is assumed to be linear and inverse of the slope of X-T diagram gives shockspeed. }
	\label{fig:figure18xtdiagram}
\end{figure}

The propagation speed $U_s$ is independently measured using the X-T diagrams of near wall void fraction, as shown in Figure  \ref{fig:figure18xtdiagram}.  Comparing the estimated Us from X-T diagrams and $U_{RH}$ computed using Figure  \ref{eqn:shockspeed} shows a very good agreement. 

\begin{figure}
	\centering
	\includegraphics[scale=1.1]{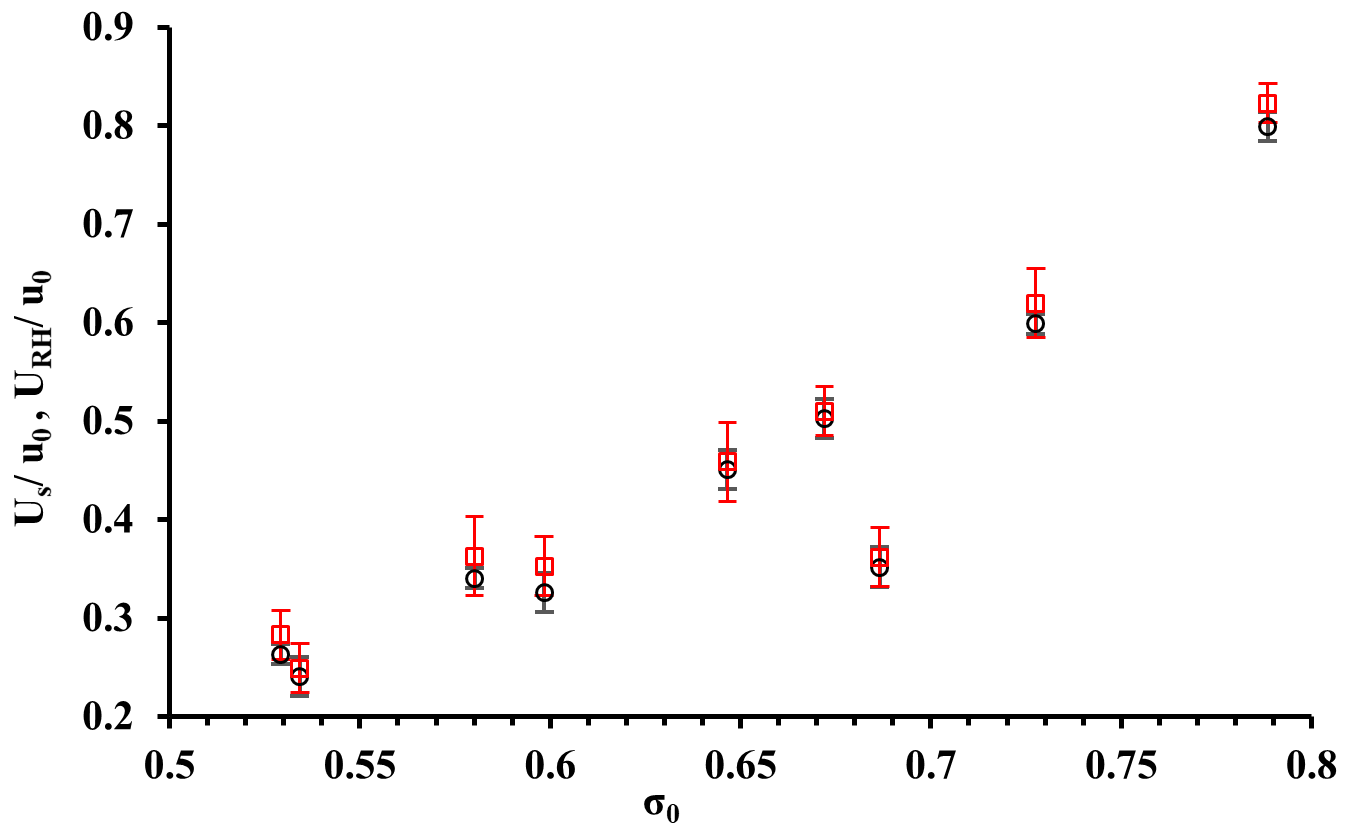}
	\caption{Comparison between computed shock-speed, $ U_{RH} ( \color{red}\square $)  and shock-speed estimated from X-T diagrams, $ U_S (\circ) $ for varying $ \sigma_{0} $ shows a good agreement between $U_S$ and $U_{RH}$.}
	\label{fig:figure19comparingus}
\end{figure}

\subsection{Speed of Sound Estimation}

The compressibility of the vapour-liquid mixture affects the periodic shedding of the cavity, evidenced by the formation of the propagating bubbly shock fronts. Quantifying the local speed of sound of the bubbly mixture helps elucidate the tendency of the flow to become supersonic at a given $\sigma_0 $.     

Due to complex nature of the rich bubbly mixtures seen in type B and type C flows, and absence of local velocity/pressure fields it is difficult to make precise experimental measurement of the local speed of sound. Instead we can estimate the local speed of sound by using simplified expressions suggested in the literature (\cite{brennen2005fundamentals}). By considering homogenously dispersed phase (V: vapour) in carrier fluid (L: water) and absence of any relative motion between the phases the governing conservation equations simplify to their single phase form. Imposing additional thermodynamic constraints leads to `the homogeneous equilibrium model’ and `the homogeneous frozen model’ for the speed of sound. The equilibrium model assumes instantaneous heat transfer between phases and mixture existing in thermodynamic equilibrium. The frozen model on the other hand considers zero heat transfer between phases (i.e. no mass transfer). The real two-phase mixture response should lie somewhere between these two extreme assumptions (\cite{brennen2005fundamentals}). If $\epsilon_L$ fraction of the liquid phase exists in thermodynamic equilibrium, then $ (1- \epsilon_{L}) $ fraction will be insulated (similarly for $ \epsilon_{V} $ fraction of the vapour phase). The general qualitative expression that can be used for both the models to estimate the speed of sound ($ c_i $) is given by: 

\begin{equation}
\frac{1}{\rho_m {c_i}^2}=\frac{\alpha}{p_i}\left[ (1-\epsilon_V) f_V+ \epsilon_V g_V\right] +\frac{1-\alpha}{{p_i}^{1+\eta}}\epsilon_L g^* {p_c}^{\eta}
\label{eqn:soundspeed}
\end{equation}

where, i=1,2 \& 3, corresponds to the pressure measured at the three static pressure taps beneath the cavity. Mixture density $ \rho_m = \alpha \rho_V + (1-\alpha)\rho_L $. Quantities f and g define thermodynamic properties of respective phase. For water $ g^* $  = 1.67, $ \eta $ = 0.73, $ g_V $  = 0.91 and $ f_V $  = 0.73 (\cite{brennen2005fundamentals}). The pressure $ p_c $ ( = 22.06 MPa) is the critical pressure of water. $\epsilon_V=\epsilon_L = 1 $ corresponds to ‘homogeneous equilibrium model’ while  $\epsilon_V=\epsilon_L = 0 $ gives the ‘homogeneous frozen model’ from Figure  \ref{eqn:soundspeed}. The Mach number is then calculated using $ u_H $ as:

\begin{equation}
M_i=\frac{u_H}{c_i}
\end{equation}

\begin{figure}
	\centering
	\includegraphics[scale=0.95]{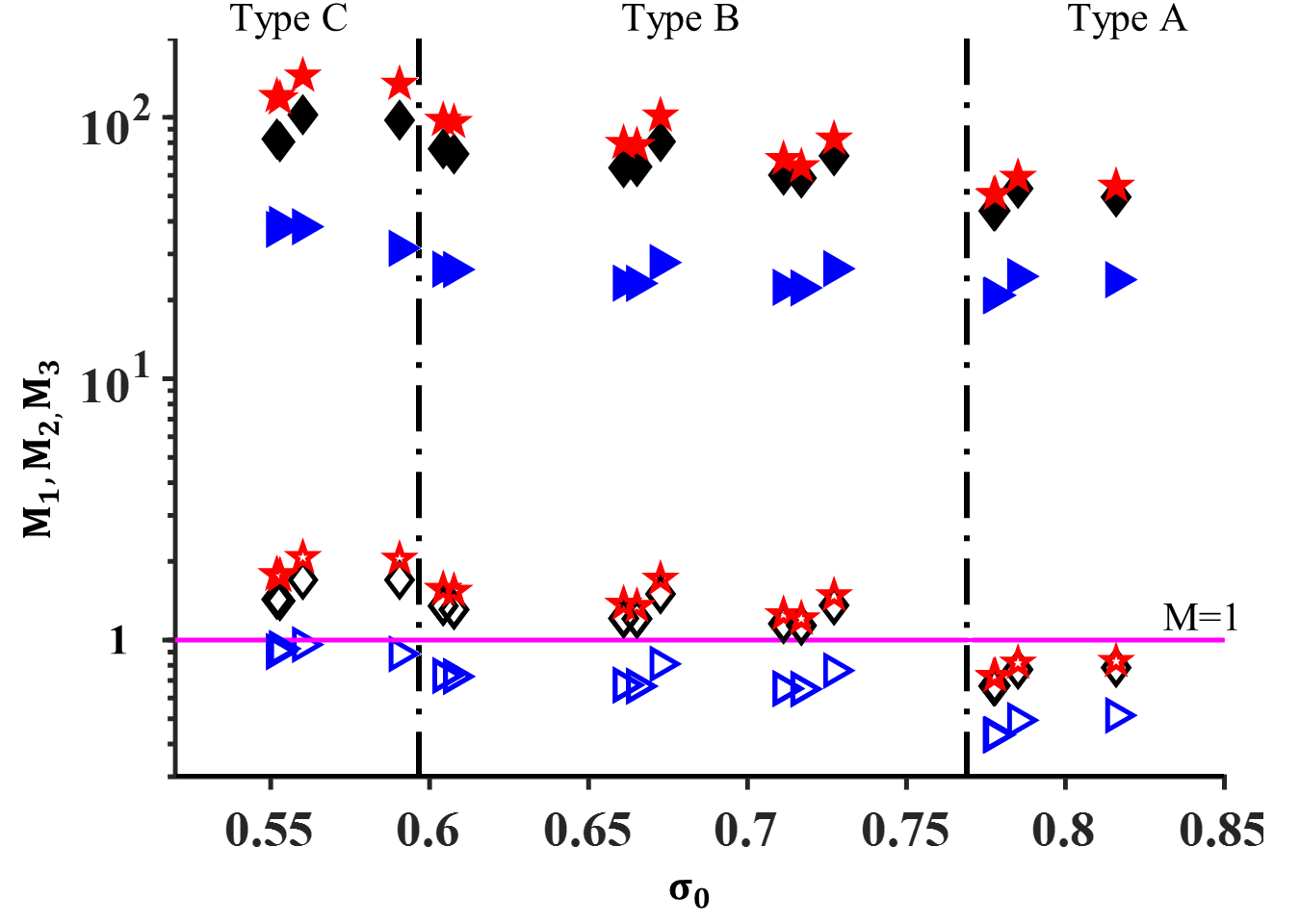}
	\caption{$ M_i $ estimated using the ‘homogeneous equilibrium model’ ($\color{red}\bigstar \color{black}\bLozenge \color{blue}\blacktriangleright $) and ‘homogeneous frozen model’ ($\color{red}\openbigstar \color{black}\bDiamond \color{blue}\rhd  $) for speed of sound;  $M_1$ ($\color{red}\bigstar \color{red}\openbigstar $ ) $M_2$ ($\color{black}\bLozenge \color{black}\bDiamond  $) \& $M_3$ ($ \color{blue}\blacktriangleright \color{blue}\rhd $).  $ u_H  = 10 m/s$. ($\color{magenta}-\color{black} $) depicts M = 1 and indicates when the flow becomes supersonic}
	\label{fig:figure20frozenvseqmmodel}
\end{figure}

Figure  \ref{fig:figure20frozenvseqmmodel} highlights that the experimentally observed onset of shockwaves (Type B shedding regime shown in Figure  \ref{fig:figure7xrtimeseries}) approximately coincides with the qualitative estimation of sound-speed using the ‘homogeneous frozen model’. The flow is seen to transition from subsonic regime at Type A to supersonic ($M>1$) at the onset of Type B regime for $ M_1$ and $M_2 $. $ M_3 $ being subsonic($<1$) can be attributed to the selection of $u_H$ instead of the local mean velocity near $p_3$ (mean flow velocity at $p_3$ is not directly measured). The ‘homogeneous equilibrium model’ is seen to consistently overestimate $ M_i $ and does not capture the subsonic to supersonic transition suggested by the formation of shock fronts (Type A to Type B transition region). Thus, this suggests that the actual speed of sound is closer to that given by the ‘homogeneous frozen model’.

\begin{figure}
	\centering
	\includegraphics[scale=0.95]{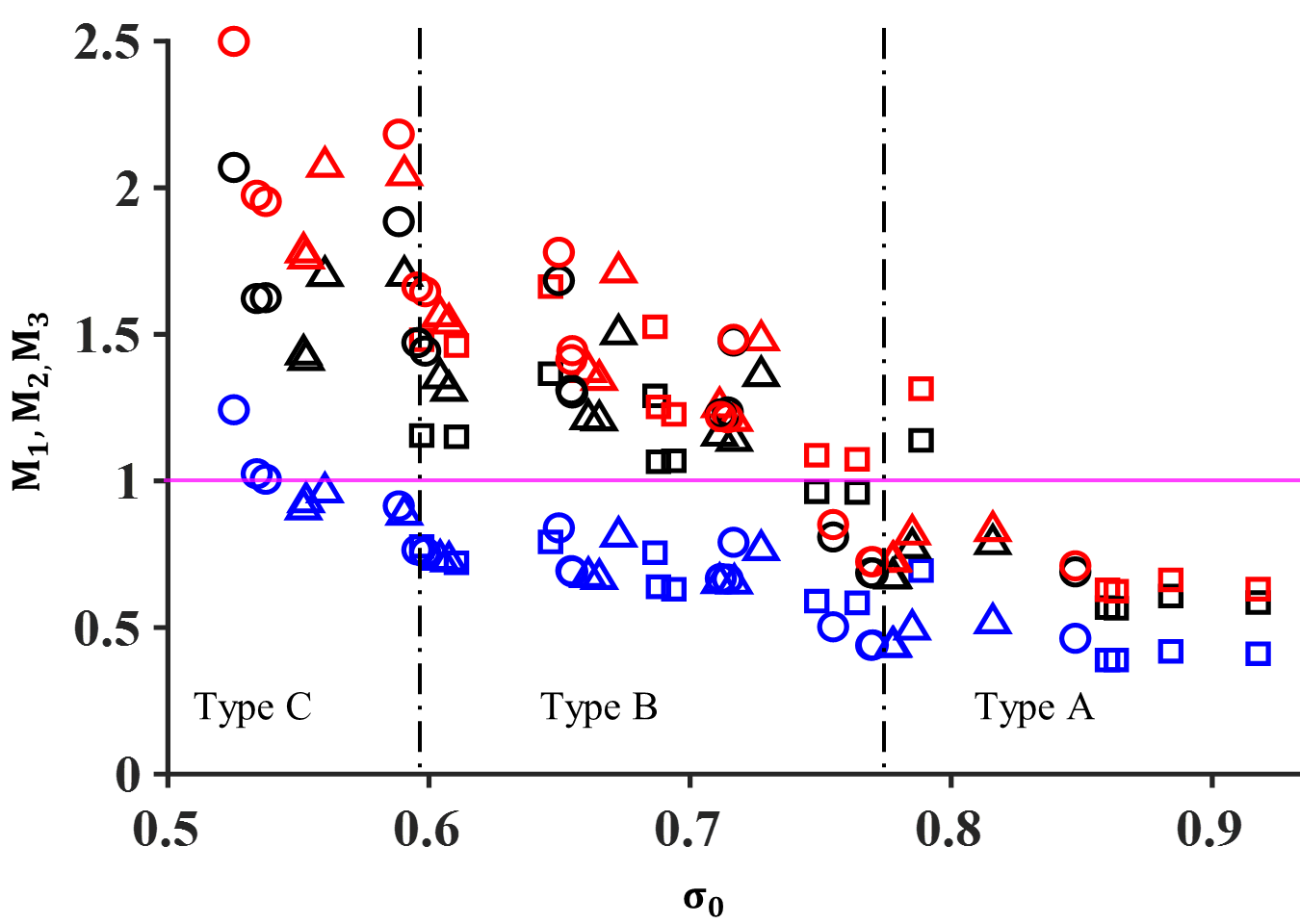}
	\caption{The variation of $ M_i $ with decreasing $ \sigma_{0} $ across three different Re estimated using ‘homogeneous frozen model’; $ u_H  = 8 m/s (\square \color{red}\square \color{blue}\square\color{black}), \color{black}10 m/s (\triangle \color{red}\triangle \color{blue}\triangle\color{black}) \color{black}$ \& $12 m/s (\circ \color{red}\circ \color{blue}\circ\color{black}); \color{black} M_1 (\square \triangle \circ),\color{black} M_2 (\color{red}\square \triangle \circ \color{black}) \ \& \  \color{black} M_3 (\color{blue} \square \triangle \circ \color{black})$. (\color{magenta}-\color{black}) depicts M = 1.}
	\label{fig:figure21m1m2m3plots}
\end{figure}

The Mach number values estimated using the ‘homogenous frozen model’ for the three Re are shown in Figure  \ref{fig:figure21m1m2m3plots}.  It is seen that $ M_i $ decreases monotonically with increasing $ \sigma_{0}$.  $M_1$, $M_2$ and $M_3$  do not have a clear Re dependency.

\section{Conclusions}

In the current study on a backward facing step we have identified three cavitating regimes observable under the x-ray densitometry and high speed cinematographic system. Throughout the study we determine the effect of change in Re on various cavity properties. 

Mean cavity lengths have a Re dependency as seen in Figure \ref{fig:figure5cavitylength}.  During a shedding cycle the stronger spanwise vortices incept and grow in size as they convect away from the separation point. The vortex-pair interaction dominates the cavity filling part of the shedding cycle as visualized in Figure \ref{fig:figure6snip}. An adverse flow front (shockwave) moving towards the step eventually diminishes the cavity (Figure \ref{fig:figure7xrtimeseries}(g)-(i)). The shedding frequency decreases monotonically with decreasing $ \sigma_{0}$ and does not have Re dependence. The mean void fraction values in the cavity are comparable across the three Re.

Static pressure $p_1$, $p_2$ and $p_3$ inside/around the cavity are measured as highlighted in Table \ref{tab:my-table}. The local pressure first drops and then recovers between $p_1$ and $p_3$. Looking at $Cp_1$ in Figure  \ref{fig:figure15sigmacvssigma0} shows that the cavity is not homogenously filled with vapour. As $ \sigma_{0}$ is dropped (Type C and below) $p_1$ is seen to approach vapour pressure.

The dynamic pressure (\textit{dp}) gives insight into the pressure fluctuations induced due to convecting shock-fronts. The shock speed computed using a simple 1-D Rankine Hugoniot jump condition lines well with the shock-speed estimated through the X-T diagram. The simple 1D assumption holds true even for shockwaves in the current complex flow system.

Finally a simplified expression is used to qualitatively estimate the local speed of sound and corresponding Mach number are discussed. The ‘homogeneous frozen model’ appears to be consistent with the observed formation of shock fronts with decreasing cavitation number.

\section{Acknowledgements}

Current study has been funded by Office of Naval Research as a part of Multidisciplinary University Research Initiative(MURI) titled `Predicting turbulent multi-phase flows with high fidelity' (Grant: N00014-17-1-2676) under program manager Dr. Ki-Han Kim.


\nocite{*}
\bibliography{allreferences}

\begin{thebibliography}{24}
\expandafter\ifx\csname natexlab\endcsname\relax\def\natexlab#1{#1}\fi
\providecommand{\bibinfo}[2]{#2}
\ifx\xfnm\relax \def\xfnm[#1]{\unskip,\space#1}\fi
\bibitem[{Aeschlimann et~al.(2011)Aeschlimann, Barre \&
  Legoupil}]{aeschlimann2011x}
\bibinfo{author}{Aeschlimann, V.}, \bibinfo{author}{Barre, S.}, \&
  \bibinfo{author}{Legoupil, S.} (\bibinfo{year}{2011}).
\newblock \bibinfo{title}{X-ray attenuation measurements in a cavitating mixing
  layer for instantaneous two-dimensional void ratio determination}.
\newblock {\it \bibinfo{journal}{Physics of Fluids}\/},  {\it
  \bibinfo{volume}{23}\/}, \bibinfo{pages}{055101}.
\bibitem[{Barbaca et~al.(2019)Barbaca, Pearce, Ganesh, Ceccio \&
  Brandner}]{barbaca2019unsteady}
\bibinfo{author}{Barbaca, L.}, \bibinfo{author}{Pearce, B.~W.},
  \bibinfo{author}{Ganesh, H.}, \bibinfo{author}{Ceccio, S.~L.}, \&
  \bibinfo{author}{Brandner, P.~A.} (\bibinfo{year}{2019}).
\newblock \bibinfo{title}{On the unsteady behaviour of cavity flow over a
  two-dimensional wall-mounted fence}.
\newblock {\it \bibinfo{journal}{Journal of Fluid Mechanics}\/},  {\it
  \bibinfo{volume}{874}\/}, \bibinfo{pages}{483--525}.
\bibitem[{Belahadji et~al.(1995)Belahadji, Franc \&
  Michel}]{belahadji1995cavitation}
\bibinfo{author}{Belahadji, B.}, \bibinfo{author}{Franc, J.-P.}, \&
  \bibinfo{author}{Michel, J.-M.} (\bibinfo{year}{1995}).
\newblock \bibinfo{title}{Cavitation in the rotational structures of a
  turbulent wake}.
\newblock {\it \bibinfo{journal}{Journal of Fluid Mechanics}\/},  {\it
  \bibinfo{volume}{287}\/}, \bibinfo{pages}{383--403}.
\bibitem[{Bernal \& Roshko(1986)}]{bernal1986streamwise}
\bibinfo{author}{Bernal, L.}, \& \bibinfo{author}{Roshko, A.}
  (\bibinfo{year}{1986}).
\newblock \bibinfo{title}{Streamwise vortex structure in plane mixing layers}.
\newblock {\it \bibinfo{journal}{Journal of Fluid Mechanics}\/},  {\it
  \bibinfo{volume}{170}\/}, \bibinfo{pages}{499--525}.
\bibitem[{Biswas et~al.(2004)Biswas, Breuer \& Durst}]{biswas2004backward}
\bibinfo{author}{Biswas, G.}, \bibinfo{author}{Breuer, M.}, \&
  \bibinfo{author}{Durst, F.} (\bibinfo{year}{2004}).
\newblock \bibinfo{title}{Backward-facing step flows for various expansion
  ratios at low and moderate reynolds numbers}.
\newblock {\it \bibinfo{journal}{J. Fluids Eng.}\/},  {\it
  \bibinfo{volume}{126}\/}, \bibinfo{pages}{362--374}.
\bibitem[{Bradshaw \& Wong(1972)}]{bradshaw1972reattachment}
\bibinfo{author}{Bradshaw, P.}, \& \bibinfo{author}{Wong, F.}
  (\bibinfo{year}{1972}).
\newblock \bibinfo{title}{The reattachment and relaxation of a turbulent shear
  layer}.
\newblock {\it \bibinfo{journal}{Journal of Fluid Mechanics}\/},  {\it
  \bibinfo{volume}{52}\/}, \bibinfo{pages}{113--135}.
\bibitem[{Brennen \& Brennen(2005)}]{brennen2005fundamentals}
\bibinfo{author}{Brennen, C.~E.}, \& \bibinfo{author}{Brennen, C.~E.}
  (\bibinfo{year}{2005}).
\newblock {\it \bibinfo{title}{Fundamentals of multiphase flow}\/}.
\newblock \bibinfo{publisher}{Cambridge university press}.
\bibitem[{Callenaere et~al.(2001)Callenaere, Franc, Michel \&
  Riondet}]{callenaere2001cavitation}
\bibinfo{author}{Callenaere, M.}, \bibinfo{author}{Franc, J.-P.},
  \bibinfo{author}{Michel, J.-M.}, \& \bibinfo{author}{Riondet, M.}
  (\bibinfo{year}{2001}).
\newblock \bibinfo{title}{The cavitation instability induced by the development
  of a re-entrant jet}.
\newblock {\it \bibinfo{journal}{Journal of Fluid Mechanics}\/},  {\it
  \bibinfo{volume}{444}\/}, \bibinfo{pages}{223--256}.
\bibitem[{Driver et~al.(1987)Driver, Seegmiller \& Marvin}]{driver1987time}
\bibinfo{author}{Driver, D.~M.}, \bibinfo{author}{Seegmiller, H.~L.}, \&
  \bibinfo{author}{Marvin, J.~G.} (\bibinfo{year}{1987}).
\newblock \bibinfo{title}{Time-dependent behavior of a reattaching shear
  layer}.
\newblock {\it \bibinfo{journal}{AIAA journal}\/},  {\it
  \bibinfo{volume}{25}\/}, \bibinfo{pages}{914--919}.
\bibitem[{Eaton \& Johnston(1981)}]{eaton1981review}
\bibinfo{author}{Eaton, J.}, \& \bibinfo{author}{Johnston, J.}
  (\bibinfo{year}{1981}).
\newblock \bibinfo{title}{A review of research on subsonic turbulent flow
  reattachment}.
\newblock {\it \bibinfo{journal}{AIAA journal}\/},  {\it
  \bibinfo{volume}{19}\/}, \bibinfo{pages}{1093--1100}.
\bibitem[{Franc \& Michel(1983)}]{franc1983two}
\bibinfo{author}{Franc, J.-P.}, \& \bibinfo{author}{Michel, J.-M.}
  (\bibinfo{year}{1983}).
\newblock \bibinfo{title}{Two-dimensional rotational structures in the
  cavitating turbulent wake of a wedge}.
\bibitem[{Ganesh et~al.(2016)Ganesh, M{\"a}kiharju \&
  Ceccio}]{ganesh2016bubbly}
\bibinfo{author}{Ganesh, H.}, \bibinfo{author}{M{\"a}kiharju, S.~A.}, \&
  \bibinfo{author}{Ceccio, S.~L.} (\bibinfo{year}{2016}).
\newblock \bibinfo{title}{Bubbly shock propagation as a mechanism for
  sheet-to-cloud transition of partial cavities}.
\newblock {\it \bibinfo{journal}{Journal of Fluid Mechanics}\/},  {\it
  \bibinfo{volume}{802}\/}, \bibinfo{pages}{37--78}.
\bibitem[{Hasan(1992)}]{hasan1992flow}
\bibinfo{author}{Hasan, M.} (\bibinfo{year}{1992}).
\newblock \bibinfo{title}{The flow over a backward-facing step under controlled
  perturbation: laminar separation}.
\newblock {\it \bibinfo{journal}{Journal of Fluid Mechanics}\/},  {\it
  \bibinfo{volume}{238}\/}, \bibinfo{pages}{73--96}.
\bibitem[{Iyer \& Ceccio(2002)}]{iyer2002influence}
\bibinfo{author}{Iyer, C.~O.}, \& \bibinfo{author}{Ceccio, S.~L.}
  (\bibinfo{year}{2002}).
\newblock \bibinfo{title}{The influence of developed cavitation on the flow of
  a turbulent shear layer}.
\newblock {\it \bibinfo{journal}{Physics of fluids}\/},  {\it
  \bibinfo{volume}{14}\/}, \bibinfo{pages}{3414--3431}.
\bibitem[{Jovic \& Driver(1995)}]{jovic1995reynolds}
\bibinfo{author}{Jovic, S.}, \& \bibinfo{author}{Driver, D.}
  (\bibinfo{year}{1995}).
\newblock \bibinfo{title}{Reynolds number effect on the skin friction in
  separated flows behind a backward-facing step}.
\newblock {\it \bibinfo{journal}{Experiments in Fluids}\/},  {\it
  \bibinfo{volume}{18}\/}, \bibinfo{pages}{464--467}.
\bibitem[{Katz \& O’hern(1986)}]{katz1986cavitation}
\bibinfo{author}{Katz, J.}, \& \bibinfo{author}{O’hern, T.}
  (\bibinfo{year}{1986}).
\newblock \bibinfo{title}{Cavitation in large scale shear flows}, .
\bibitem[{O'hern(1990)}]{o1990experimental}
\bibinfo{author}{O'hern, T.} (\bibinfo{year}{1990}).
\newblock \bibinfo{title}{An experimental investigation of turbulent shear flow
  cavitation}.
\newblock {\it \bibinfo{journal}{Journal of fluid mechanics}\/},  {\it
  \bibinfo{volume}{215}\/}, \bibinfo{pages}{365--391}.
\bibitem[{Scarano et~al.(1999)Scarano, Benocci \&
  Riethmuller}]{scarano1999pattern}
\bibinfo{author}{Scarano, F.}, \bibinfo{author}{Benocci, C.}, \&
  \bibinfo{author}{Riethmuller, M.} (\bibinfo{year}{1999}).
\newblock \bibinfo{title}{Pattern recognition analysis of the turbulent flow
  past a backward facing step}.
\newblock {\it \bibinfo{journal}{Physics of Fluids}\/},  {\it
  \bibinfo{volume}{11}\/}, \bibinfo{pages}{3808--3818}.
\bibitem[{Spazzini et~al.(2001)Spazzini, Iuso, Onorato, Zurlo \&
  Di~Cicca}]{spazzini2001unsteady}
\bibinfo{author}{Spazzini, P.}, \bibinfo{author}{Iuso, G.},
  \bibinfo{author}{Onorato, M.}, \bibinfo{author}{Zurlo, N.}, \&
  \bibinfo{author}{Di~Cicca, G.} (\bibinfo{year}{2001}).
\newblock \bibinfo{title}{Unsteady behavior of back-facing step flow}.
\newblock {\it \bibinfo{journal}{Experiments in fluids}\/},  {\it
  \bibinfo{volume}{30}\/}, \bibinfo{pages}{551--561}.
\bibitem[{Troutt et~al.(1984)Troutt, Scheelke \& Norman}]{troutt1984organized}
\bibinfo{author}{Troutt, T.}, \bibinfo{author}{Scheelke, B.}, \&
  \bibinfo{author}{Norman, T.} (\bibinfo{year}{1984}).
\newblock \bibinfo{title}{Organized structures in a reattaching separated flow
  field}.
\newblock {\it \bibinfo{journal}{Journal of Fluid Mechanics}\/},  {\it
  \bibinfo{volume}{143}\/}, \bibinfo{pages}{413--427}.
\bibitem[{Wee et~al.(2004)Wee, Yi, Annaswamy \& Ghoniem}]{wee2004self}
\bibinfo{author}{Wee, D.}, \bibinfo{author}{Yi, T.},
  \bibinfo{author}{Annaswamy, A.}, \& \bibinfo{author}{Ghoniem, A.~F.}
  (\bibinfo{year}{2004}).
\newblock \bibinfo{title}{Self-sustained oscillations and vortex shedding in
  backward-facing step flows: Simulation and linear instability analysis}.
\newblock {\it \bibinfo{journal}{Physics of Fluids}\/},  {\it
  \bibinfo{volume}{16}\/}, \bibinfo{pages}{3361--3373}.
\bibitem[{Wu(2019)}]{wu2019bubbly}
\bibinfo{author}{Wu, J.} (\bibinfo{year}{2019}).
\newblock {\it \bibinfo{title}{Bubbly Shocks in Separated Cavitating Flows}\/}.
\newblock Ph.D. thesis.
\bibitem[{Wu et~al.(2019)Wu, Ganesh \& Ceccio}]{wu2019multimodal}
\bibinfo{author}{Wu, J.}, \bibinfo{author}{Ganesh, H.}, \&
  \bibinfo{author}{Ceccio, S.} (\bibinfo{year}{2019}).
\newblock \bibinfo{title}{Multimodal partial cavity shedding on a
  two-dimensional hydrofoil and its relation to the presence of bubbly shocks}.
\newblock {\it \bibinfo{journal}{Experiments in Fluids}\/},  {\it
  \bibinfo{volume}{60}\/}, \bibinfo{pages}{66}.
\bibitem[{Young \& Holl(1966)}]{young1966effects}
\bibinfo{author}{Young, J.}, \& \bibinfo{author}{Holl, J.~W.}
  (\bibinfo{year}{1966}).
\newblock \bibinfo{title}{Effects of cavitation on periodic wakes behind
  symmetric wedges}, .

\end{thebibliography}

\end{document}